\documentclass[11pt,a4paper]{article}
\usepackage{jcappub}
\usepackage{amsmath}
\usepackage{amssymb}
\usepackage{amsfonts}
\usepackage{graphicx,bm}
\usepackage{dcolumn}
\usepackage{color,amsxtra}
\usepackage{epsf}
\usepackage{enumerate}
\usepackage{hhline}
\usepackage{array}
\usepackage{tabularx}
\usepackage{subfigure}
\usepackage{appendix}%Include appendix
\usepackage[colorlinks=true]{hyperref}
\newcommand{\be}{\begin{equation}}
\newcommand{\ee}{\end{equation}}
\newcommand{\bea}{\begin{eqnarray}}
\newcommand{\eea}{\end{eqnarray}}
\newcommand{\beaa}{\begin{eqnarray*}}
\newcommand{\eeaa}{\end{eqnarray*}}

\newcommand{\nn}{\nonumber \\}
\newcommand{\e}{\mathrm{e}}

\newcommand{\Eqn}[1]{&\hspace{-0.2em}#1\hspace{-0.2em}&}
\def\be{\begin{equation}}
\def\ee{\end{equation}}
\def\bea{\begin{eqnarray}}
\def\eea{\end{eqnarray}}

\def\nn{\nonumber \\}
\def\e{\mathrm{e}}

\begin{document}

\title{%Inflation in $f(R,\xi)$ gravity: analyzing
Spotting deviations from $R^2$ inflation}
%Deviations from $R^2$-inflation in  $f(R,\xi)$ gravity theories}

\author{\'Alvaro de la Cruz-Dombriz$^{a}$\ \footnote{alvaro.delacruzdombriz[at]uct.ac.za}, Emilio Elizalde$^{b}$\ \footnote{elizalde[at]ieec.uab.es}, Sergei D. Odintsov$^{b,c}$\ 
\footnote{odintsov[at]ieec.uab.es}, Diego S\'aez-G\'omez$^{d}$\ \footnote{dsgomez[at]fc.ul.pt}}
%\email{}\affiliation{}
\affiliation{${}^a$Astrophysics, Cosmology and Gravity Centre (ACGC), Department of Mathematics and Applied Mathematics, University of Cape Town, Rondebosch 7701, Cape Town, South Africa \\
${}^b$ Institut de Ci\`{e}ncies de l'Espai, ICE/CSIC-IEEC,
Campus UAB, Carrer de Can Magrans s/n,
08193 Bellaterra (Barcelona), Spain\\
${}^c$Instituci\'{o} Catalana de Recerca i Estudis Avan\c{c}ats
(ICREA), Barcelona, Spain\\
${}^d$ Instituto de Astrof\'isica e Ci\^encias do Espa\c{c}o, Faculdade de Ci\^encias da Universidade de Lisboa, Edif\'icio C8, Campo Grande, P-1749-016
Lisbon, Portugal}
%\date{}

%%%%%%%%%%%%%%%%%%%%%
% Abstract
%%%%%%%%%%%%%%%%%%%%%
\abstract{
We discuss the soundness of inflationary scenarios in theories beyond the Starobinsky
model, namely a class of theories described by arbitrary functions of the Ricci scalar and the $K$-essence field.
%
%, the trace of the energy-momentum tensor and the contraction of the Ricci tensor with the energy-momentum tensor.
We discuss the pathologies associated with higher-order equations of motion which will be shown to constrain
 the stability of this class of theories. We provide a general framework to calculate the
 slow-roll parameters and the corresponding mappings to the theory parameters.
For paradigmatic gravitational models within the class of theories under consideration we illustrate
the power of the Planck/Bicep2 latest results to constrain such gravitational Lagrangians. Finally, bounds for potential deviations
from Starobinsky-like inflation are derived.}

\maketitle

%\def\thesection{\Roman{section}}
%\def\theequation{\Roman{section}.\arabic{equation}}

%%%%%%%%%%%%%%%%%%%%%%%%%%%
%%%  Sec. I
%%%%%%%%%%%%%%%%%%%%%%%%%%%
\section{Introduction}

A key question to be addressed in gravitational physics and suggested extensions to General Relativity (GR)  %able to address some of the open problems in Cosmology
resides in the coupling of gravity and matter fields. In this respect, the assumption of the Equivalence Principle and consequently,  the minimal coupling between matter and geometry  as dictated by Einsteinian gravity, although supported by both astrophysical tests and laboratory experiments, may suffer from violations %of the minimal coupling violation
in spacetime scales  where experiments are neither conclusive enough nor have never been performed.
Indeed, the Einstein Equivalence Principle is a fundamental milestone of the GR theory. While this principle has been thoroughly tested with standard matter, the question of its validity in the dark sector remains open.
% Motivation
%
The existing literature has considered  various forms of non-minimal couplings \cite{Clifton:2011jh} 
such as those provided by scalar-tensor theories \cite{Scalar-tensor proposals}, vector-tensor theories \cite{Vector proposals},
%(see also \cite{Clifton:2011jh} for a thorough review on different proposals for non-minimal couplings)
different couplings between matter and geometry \cite{matter-geometry couplings},
and even couplings of
%
% A recently proposed departing point consists of suggesting the coupling of
a function of the Ricci scalar $R$ to the matter Lagrangian
%a proposal which has produced numerous studies for gravitational and cosmological issues on the subject
({\it c.f.} \cite{Koivisto 2013, varia-fR-Lagrangian, Dombriz-fRTRT, Odintsov:2013iba, Haghani:2013oma} and references therein).
%
%
%
%----------------------------------------------
% f(R,T) theories
A recent line of research has also considered Lagrangians which depend both on the Ricci curvature %$R$
and on the trace of the energy-momentum tensor $T=T^{\mu}_{\;\;\mu}$. Originally introduced in
\cite{Poplawski:2006ey} and later in \cite{fRTpaper}, %by Harko {\it et al.}
some cosmological aspects have been already explored for these kinds of theories, such as the reconstruction of cosmological solutions \cite{Varia_fRT, stephaneseul3}
and other issues on energy conditions, thermodynamics and singularities~\cite{flavio, thermo1, Harko_August_2014}.
%
% Also the energy conditions have been analyzed in Ref.~\cite{flavio}. The thermodynamics of Friedmann-Lema\^itre-Robertson-Walker (FLRW) spacetimes has been studied in Ref.~\cite{thermo1}. More recently the possibility of irreversible matter creation processes and the possibility of the occurrence of future singularities were addressed in \cite{Harko_August_2014} and  \cite{juliano} respectively.
%
%
However, theories with non standard couplings between the geometry and the matter Lagrangian % (see \cite{Nesseris:2008mq})
usually fail to conserve the energy-momentum tensor, which implies a stringent shortcoming for their viability \cite{Diego-Alvaro-PRD2013}.
%
% f(R,T,RmunuTmunu)
Another kind of theory under scrutiny in the recent years also included terms of the form $R_{\mu\nu}T^{\mu\nu}$  \cite{Odintsov:2013iba,  Odintsov_HL, Dombriz-fRTRT}.
Finally, a recent proposal in \cite{Julien} has claimed that  in the case of a violation of the Equivalence Principle, data favour violations through coupling strengths showing opposite signs for ordinary and dark matter, although the analysis therein does not show any significant deviations from GR.
\\

These different types of couplings, if intended to depict a well-founded theory describing the gravitational interaction, must be free of undesirable instabilities
such as the appearance of ghost-like modes and the exponential growth of perturbations around well-established spacetime backgrounds, among others.
For instance, %an undesirable instability is the so-called Dolgov-Kawasaki instability which appears when at least one extra degree of freedom of the theory behaves as a ghost and therefore this mode would act to destabilise the theory with no stable ground state.
the avoidance of the Dolgov-Kawasaki instability has been developed to constrain the extensively studied $f(R)$ gravity with minimal \cite{DK, Sergei_DK, DK-HuSawicki, Faraoni2007, Cognola:2007zu} and non-minimal couplings of the curvature with matter \cite{Dolgov-Kawasaki-non-minimal}.
%
% Authors in  have recently addressed this instability issue for $f(R,T, R_{\mu\nu}T^{\mu\nu})$ theories.
%
%
Another important requirement usually demanded to extended theories consists of the avoidance of the Ostrogradski instability, % i.e., the fact that
%Ostrogradski's theorem:
% a linear instability
which appears in Hamiltonians associated with Lagrangians depending on more than one time derivative non-degenerately \cite{Ostrogradski, Woody}. As a consequence, such Hamiltonians
turn out not to be bounded from below and well-defined vacuum states are absent.
However, theories with higher-order equations of motion can be sensible provided they are regarded in the context of effective field theories, where operators leading to the higher-order equations of motion comprise the first terms of some expansion whose adequate resummation might give rise to well-behaved theories.
%(this is the situation for instance when one integrates heavy degrees of freedom out).
Another possibility to make sense of theories with higher-order equations of motion is to remove the undesired unstable degrees of freedom from the physical spectrum of the theory or, at the classical level, to constrain the physically allowed set of boundary conditions. However, one needs to make sure that such a procedure does not get spoiled by either time-evolution or coupling to other fields. This approach was followed in \cite{Jimenez:2012ak} for the case of the degenerate Pais-Uhlenbeck field.
A widely accepted way of circumventing the Ostrogradski instability when considering scalar-tensor theories of gravitation, consists of requiring the Euler-Lagrange equations to be second order even if higher-order derivatives are present in the action. Following this line of reasoning, Horndeski's theorem \cite{Horndeski} provides the most general Lagrangian density for a scalar-tensor theory which guarantees second-order Euler-Lagrange equations. %, being for instance Galileons theories \cite{Galileons} subsets  of Horndeski-like theories.
Nonetheless, recent proposals have ensured second order equations of motion and hence the absence of Ostrogradski ghost degrees of freedom
for theories which do not fall under the form of Horndeski-like theories  ({\it c.f.} \cite{Piazza, Zuma, MultiScalar, Non-local, NonLinearHorndeski} and references therein for recent proposals).
\\

In this paper we shall focus on theories presenting a general coupling through an arbitrary function between the Ricci scalar curvature $R$ and a 
calar field given by a $K$-essence model, the latter including in principle both a kinetic and a potential term.
% f(R) theories
Arbitrary theories of the Ricci scalar only, widely known as $f(R)$ theories might be thought of as the only local, metric-based and generally coordinate invariant and stable modifications of gravity \cite{Woody, Non-local}.
When rearranged in the scalar-tensor equivalent picture by the introduction of an auxiliary scalar field, those theories prove to be free from Ostrogradski instability\footnote{Gravitational Lagrangians which are arbitrary functions of the Lovelock invariant can also be proved to be free of Ostrogradski instability \cite{Bueno-UAM}.}. Moreover, both viability and stability conditions for $f(R)$ theories have been widely studied and guarantee the attractive character, the aforementioned avoidance of the Dolgov-Kawasaki instability, and the agreement with solar system tests and evolution of geodesics \cite{JCAP-Alvaro}.
$f(R)$ theories have in fact proved extremely successful in describing the cosmological evolution, being able to
provide a complete picture of the behaviour of the Universe  ({\it c.f.} \cite{delaCruzDombriz:2006fj,  Cognola:2007zu, Nojiri:2007cq}) %) and references therein)
accounting for the dark matter component \cite{Cembranos:2008gj, Capozziello:2012ie}, the growth of large-scale structures \cite{perturbations}
and to provide a mechanism for cosmological inflation through, for instance, the Starobinsky inflationary model \cite{staro, Planck-Inflation}, which is constructed with an extra term to the Einstein-Hilbert Lagrangian of the form $f(R)=R^2$. Other recent proposals have been scrutinised in \cite{applications_staro,Bamba:2014daa,Bamba:2014wda}.
One is then tempted to consider general theories of the Ricci scalar with some kind of coupling to an extra scalar field in order to observe potential effects distinguishable from
the standard $f(R)$ counterpart predictions. In particular, the scalar field might be interpreted either as an effective field theory  remnant
from an underlying theory or the inflaton field as it shall be the case below.
Namely, in the following we shall focus our attention on the inflationary epoch and our goal throughout the paper will be twofold: on the one hand, we shall study how to reconstruct a given inflationary Hubble parameter for classes of gravitational models of the sort described above; on the other hand, we shall determine how deviations from the Starobinsky inflationary scenario when an extra scalar field is introduced can be identified, or in other words, how certain the claim  that Starobinsky inflation is in fact the model behind the observed slow-roll parameters may be. \\

Slow-roll inflation is usually described by a single scalar field, where the desirable features can be achieved by the appropriate scalar potential \cite{Mukhanov,Elizalde:2008yf}. However, while considering several scalar fields, slow-roll inflation can be also reproduced. In this sense, modifications studied herein will be analysed in the Einstein frame, where our theory turns out precisely to be interpreted as a multifield inflationary model. As well known, in the multifield inflation paradigm the conservation of the curvature perturbation is not ensured at scales beyond the horizon, but eventual conservation depends on the evolution of the several scalar fields, inducing isocurvature modes ({\it c.f.} Refs.\cite{Sasaki:1995aw}-\cite{Kaiser:2013sna}). However, the spectral index for the adiabatic modes can be analytically obtained \cite{Sasaki:1995aw,GarciaBellido:1995qq}. In addition, even if isocurvature modes are created, they may become adiabatic when the Universe reaches thermal equilibrium \cite{Weinberg:2004kf}. In addition, tools for calculating and constraining multifield inflation has been developed, being so far in good agreement with the observational constraints \cite{Kaiser:2013sna}. In the following we shall analyse some modifications of Starobinsky $R^2$ inflationary model which imply the emergence of a multifield picture, where the spectral index and the scalar-to-tensor ratio can be studied.\\

% {\bf Which modifications to consider?}
The modifications to be considered in the following sections are supported by several theoretical motivations which appear when the nature of GR is understandably demoted to an effective theory. Thus, 
whenever general covariant terms in a gravitational low-energy expansion are considered, powers of the Ricci scalar as well as contraction of both Ricci and Riemman tensors appear in the gravitational sector \cite{Donoghue}, a phenomenon that for instance explains the interest and theoretical support of scalar-tensor $f(R)$ theories
as well as other extended theories of gravity ({\it c.f.}~\cite{Nojiri:2007cq} for extensive reviews on the subject). Moreover, the natural emergence of scalar fields  in string-inspired theories is also a well-known phenomenon \cite{ScalarFieldsStrings} which serves to motivate the addition of scalar fields, for instance in the form of $K$-essence Lagrangians in the calculations to follow. 
Finally a desirable attribute in the theories suggested below, is indeed related to the absence of Ostrogradski's instability, so we shall justify how the emerging non-minimal couplings do not spoil such a worthwhile attribute. Consequently, the effective gravitational actions with which we shall deal in the bulk of the paper, could be thought of as including both geometric corrections in the Einstein-Hilbert  Lagrangian and an additional scalar field.
\\

The paper is structured as follows: in Sec. \ref{Sec:f(R,xi)} we  present some generalities of the theories under consideration. %, paying special attention to the theoretical limitations imposed by the fact of considering the energy-momentum tensor at the level of the gravitational action.
Therein we  include the multi-scalar representation, which allows us to identify the potential instabilities in a transparent way. Then, we  perform the conformal transformation to the Einstein frame in order to express
the gravitational sector in a more convenient way. At the end of the Section we write the field equations in a cosmological spatially flat Robertson-Walker geometry that will be used in the following.
 In Sec. \ref{Sec:Slow-roll} we will focus on the inflationary  slow-roll paradigm and we recast the inflationary slow-roll parameters in terms of the Hubble parameter and its derivatives, which allows us to express such parameters in terms of the cosmological evolution as provided by the  gravitational modelss considered. Immediately after that, in Sec. \ref{Sec:Reconstruction}
 we will apply the technique developed in the previous Section to two different parameterisations for the Hubble rate at the inflationary epoch. We illustrate the results using two paradigmatic gravitational models, one in Sec. \ref{SubSec:ModelA} with a multiplicative coupling between the curvature and the $K$-essence field and a second one in Sec. \ref{SubSec:ModelB}
 with  the addition of an arbitrary function of the Ricci curvature and another function of the $K$-essence scalar field. In Sec. \ref{Sec:Deviations} we perform a thorough study on two possible functional deviations from the standard Starobinsky model as originated by arbitrary functions of a $K$-essence scalar field.  We shall perform explicit calculations for minimal and non-minimal couplings between the Starobinsky model  and $K$-essence in Secs. ~\ref{Subsection:5.1} and \ref{Subsection:5.2} respectively and provide the corresponding slow-roll parameters.
Finally, Sec. \ref{Sec:Conclusions} contains the main conclusions of the paper. % and an outlook.

%
% \label{Sec:f(R,xi)}
%\label{Sec:Slow-roll}
%\label{Sec:Reconstruction} \label{SubSec:ModelA} \label{SubSec:ModelB}
%\label{Sec:Deviations}
%\label{Sec:Conclusions}

%%%%%%%%%%%%%%%%%%%%%%%%%%%%%%%%%%%%%%%%%%%%%%%%
%%%%%%%%%%%%%%%%%%%%%%%%%%%%%%%%%%%%%%%%%%%%%%%%%%
%
% Planck units, $(G=c=k_B =\hbar=4\pi \varepsilon_0 =1)$ will be used
% Throughout this paper, Greek indices run from 0 to 3, the symbol $\nabla$ denotes the standard covariant derivative and the signature $+,-,-,-$ is used.
%The Riemann tensor definition
%is $R^\mu_{\;\;\nu\alpha\beta}=
%\partial_\alpha\Gamma^{\mu}_{\nu\beta}
%-\partial_\beta\Gamma^{\mu}_{\nu\alpha}
%+\Gamma^{\mu}_{\sigma\alpha}\Gamma^{\sigma}_{\nu\beta}
%-\Gamma^{\mu}_{\sigma\beta}\Gamma^{\sigma}_{\nu\alpha}$.
% which coincides with the definition in has opposite sign to the one proposed in \cite{Giovannini}.}.

%%%%%%%%%%%%%%%%%%%%%%%%%%%%%%%%%%%%%%%%%%%%%%%%%%%%%%%%%%%%%%%%%%%%%%%%%%%%%%
\section{$f(R,\xi)$ gravity}
\label{Sec:f(R,xi)}
%%%%%%%%%%%%%%%%%%%%%%%%%%%%%%%%%%%%%%%%%%%%%%%%%%%%%%%%%%%%%%%%%%%%%%%%%%%%%
Let us start by introducing the general scalar-tensor model to be discussed in this manuscript. The gravitational action is expressed as follows
\be
S=\int {\rm d}^4x\sqrt{-g}\left[f(R,\xi)+2\kappa^2\mathcal{L}_m\right]\ ,
\label{1.1}
\ee
where $\kappa^2=8\pi G$, $G$ is the standard gravitational constant  and $\xi\equiv-\frac{1}{2}\partial_{\mu}\phi\partial^{\mu}\phi-V(\phi)$ that just by itself represents the Lagrangian of a canonical scalar field.
%An arbitrary funcion of $\xi$ only represents a $K$-essence field.
This kind of action generalises the usual scalar-tensor theories, such as Brans-Dicke-like or $K$-essence theories. Then, by varying the action with respect to the metric $g_{\mu\nu}$ and the scalar field $\phi$, the corresponding field equations are obtained, yielding
\bea
R_{\mu\nu}f_R-\frac{1}{2}g_{\mu\nu}f+(g_{\mu\nu}\Box-\nabla_{\mu}\nabla_{\nu})f_R-\frac{1}{2}f_{\xi}\nabla_{\mu}\phi\nabla_{\nu}\phi&=&\kappa^2T_{\mu\nu}\ , \nn
\nabla_{\mu}\left(f_{\xi}\nabla^{\mu}\phi\right)-f_{\xi}V(\phi)&=&0\ .
\label{1.2}
\eea
Here the subscripts refer to derivatives with respect to $R$ and $\xi$. As in other extensions of GR, the action (\ref{1.1}) may be expressed in terms of two auxiliary scalar fields $\{\varphi,\,\psi\}$ as follows
\bea
S=\int {\rm d}^4x\sqrt{-g}\left[\varphi R+\psi\xi-Q(\varphi,\psi)+2\kappa^2\mathcal{L}_m\right]\ .
\label{1.3}
\eea
Varying the action (\ref{1.3}) with respect to $\varphi$ and $\psi$, respectively, yields
\bea
R-\frac{\partial Q(\varphi, \psi)}{\partial \varphi}=0\ , \quad \xi-\frac{\partial Q(\varphi,\psi)}{\partial\psi}=0%\nn
\quad \rightarrow \quad \varphi=\varphi(R, \xi)\ , \quad \psi=\psi(R, \xi)\ .
\label{1.4}
\eea
where the last two expressions rely on the possibility of inverting the partial derivatives of $Q$ \cite{Dombriz-fRTRT}. Usually explicit expressions for $\varphi$ and $\psi$ are not found, but numerical ones are always reachable provided $\det\frac{\partial^2f}{\partial\chi_i\partial\chi_j}\neq0$ with $\chi_{i=1,2}=\{\varphi,\,\psi\}$ \cite{Dombriz-fRTRT}.
Thus, expressions
\bea
\varphi=f_R\ , \quad \psi=f_{\xi}\ , %\nn
\quad Q(\varphi, \psi)=f_R R+f_{\xi}\xi-f\,  %\nn \nn
\label{1.4a}
\eea
and consequently
\bea
%&\rightarrow
f(R,\xi)=\varphi R+\psi\xi-Q(\varphi, \psi)\ .
 \label{1.4b}
\eea
turn  the form of action (\ref{1.3}) back to the original one (\ref{1.1}).
Hence, we have sketched how the gravitational action (\ref{1.1}) can be expressed through (\ref{1.3}) as a scalar-tensor theory with two additional scalar fields $\{\varphi,\,\psi \}$, non-minimally coupled to the Ricci scalar and the original scalar field $\phi$, respectively. The fact that action can be rewritten in the form (\ref{1.3}) guarantees the absence of the Ostrogradski's instability for such theories\footnote{A complementary way to check the absence of Ostrogradski's instability is to consider that $f(R)$ theories are instability-free \cite{Dombriz-fRTRT}, therefore only the field $\xi$ can be problematic but an analogous process as the one exemplified in (\ref{1.3}) only for the $\xi$ field leads to the same conclusion}. Thus, once the action is written in the scalar-tensor form  (\ref{1.4b}), the latter action can be further transformed into the so-called Einstein frame, where the couplings in the curvature sector recover the usual
form of the Einstein-Hilbert action. Hence, by applying the conformal transformation:
\be
\tilde{g}_{\mu\nu}=\Omega^2 g_{\mu\nu}\ \quad \text{where} \quad \Omega^2=\varphi\ ,
\label{1.5}
\ee
the action (\ref{1.3}) yields
\be
\tilde{S}=\int {\rm d}^4x\sqrt{-\tilde{g}}\left[\frac{\tilde{R}}{2\kappa^2}-\frac{1}{2}\partial_{\mu}\tilde{\varphi}\partial^{\mu}\tilde{\varphi}+P(\tilde{\varphi}, \psi)\tilde{\xi}-\tilde{Q}(\tilde{\varphi},\psi)+\e^{-2\sqrt{\frac{2}{3}}\kappa\tilde{\varphi}}\mathcal{L}_m\right]\ ,
\label{1.6}
\ee
where the following notation has been implemented:
\be
\varphi=\e^{\sqrt{\frac{2}{3}}\kappa\tilde{\varphi}}\ ,\quad P(\tilde{\varphi}, \psi)=\frac{\e^{-2\sqrt{\frac{2}{3}}\kappa\tilde{\varphi}}}{2\kappa^2} \psi\ , \quad \tilde{Q}=\frac{\e^{-2\sqrt{\frac{2}{3}}\kappa\tilde{\varphi}}}{2\kappa^2} Q\ ,\quad \tilde{\xi}=-\frac{1}{2}\e^{\sqrt{\frac{2}{3}}\kappa\tilde{\varphi}}\partial_{\mu}\phi\partial^{\mu}\phi-V(\phi)\ .
\label{1.7}
\ee
Then, action $f(R,\xi)$, when mapped into the Einstein frame,
can be recast as a standard
Einstein-Hilbert term\footnote{Although as usual in the Einstein frame, the price to pay is the non-minimal coupling between matter and geometry as seen in the last term of  (\ref{1.6}).
} plus a $K$-essence contribution supplemented by
two auxiliary coupled scalar
fields $\{\tilde\varphi,\,\psi\}$, which do couple to each other through functions $P$ and $\tilde Q$. On the other hand, there is a coupling of the aforementioned fields with $\tilde\xi$, which appears multiplying the $P$ function in (\ref{1.6}).

\vspace{0.5cm}
In order to illustrate the power of the above procedure, let us consider a special case where calculations become simpler, $f(R, \xi)=f(R+\xi)$. In such case, $\varphi$ and $\psi$ coincide according to Eq. (\ref{1.4a}) and the action (\ref{1.3}) becomes
\be
S=\int {\rm d}^4x\sqrt{-g}\left[\varphi(R+\xi)-Q(\varphi)+2\kappa^2\mathcal{L}_m\right]\ .
\label{1.8}
\ee
There, once the conformal transformation (\ref{1.5}) is applied, one gets
\be
\tilde{S}=\int {\rm d}^4x\sqrt{-\tilde{g}}\left[\frac{\tilde{R}}{2\kappa^2}-\frac{1}{2}\partial_{\mu}\tilde{\varphi}\partial^{\mu}\tilde{\varphi}+\tilde{Q}(\tilde{\varphi})+\frac{\e^{-\sqrt{\frac{2}{3}}\kappa\tilde{\varphi}}}{2\kappa^2}\tilde{\xi}+\e^{-2\sqrt{\frac{2}{3}}\kappa\tilde{\varphi}}\mathcal{L}_m\right]\ .
\label{1.9}
\ee
Thus as can be seen from the $\tilde\xi$ definition in (\ref{1.7}), provided there is no scalar potential
$V(\phi)$ in the original action, the couplings between the scalar fields $\{\phi,\,\tilde\varphi\}$
and %with respect to
the Ricci scalar $\tilde R$ turn out to be minimal\footnote{Once again the non-minimal coupling  with matter survives, as seen in the last term of (\ref{1.9}).} when expressing the Lagrangian $f(R+\xi)$ in
the Einstein frame as we did in (\ref{1.9}).
This result has been widely studied in previous literature as it provides interesting properties and reproduces the late-time acceleration, although usually at the expense of having to suffer from a smooth phantom transition ({\it c.f.} Ref.~\cite{Elizalde:2008yf}). \\

\subsection{Cosmological Evolution}
Let us now turn back to the original equations (\ref{1.2}) and consider a spatially flat Friedmann-Lema\^itre-Robertson-Walker (FLRW) metric,
\be
{\rm d}s^2=-{\rm d}t^2+a(t)^2{\rm d}\vec{r}^{\,2}
\label{1.10}
\ee
Here, as usual $a(t)$ is the scale factor and the in the following the Hubble parameter $H \equiv \dot{a}/a$
with the dot %$\dot\ $
expressing
derivative with respect to the cosmic time $t$. Then, introducing this metric in the field equations (\ref{1.2}), the FLRW equations for the theory (\ref{1.1}) for a perfect fluid with density and pressure $\rho_m$ and $p_m$ respectively, yield %are obtained,
\bea
3f_R H^2&=&\kappa^2\rho_m+\frac{1}{2}f_{\xi}\dot{\phi}^2+\frac{1}{2}\left(Rf_{R}-f\right)-3H\dot{f_R}\,,
%\frac{\partial f_{R}}{\partial t}
\ \nn
-\left(2\dot{H}+3H^2\right)f_{R}&=&\kappa^2p_m+\frac{1}{2}\left(f-Rf_R\right)+\ddot{f_R}%\frac{\partial^2 f_R}{\partial t^2}
+2H\dot{f_R}%\frac{\partial f_R}{\partial t}
\ .
\label{1.11}
\eea
For the purposes of the paper, let us express the FLRW equations (\ref{1.11}) in terms of the number of e-folds $N\equiv \log\frac{a}{a_0}$ instead of the cosmic time $t$, which shall prove more convenient in the following. %would be more convenient.
Then, equations turn into
\bea
3f_R H^2&=&\kappa^2\rho_m+\frac{1}{2}f_{\xi}H^2\phi^{\prime2}+\frac{1}{2}\left(Rf_{R}-f\right)-3H^2 f_R^{\prime}\,,%\frac{\partial f_{R}}{\partial N}
\ \nn
-\left(2 HH^{\prime}+3H^2\right)f_{R}&=&\kappa^2p_m+\frac{1}{2}\left(f-Rf_R\right)+H^2f_{R}^{\prime\prime}%\frac{\partial^2 f_R}{\partial N^2}
+\left(2H^2+H H^{\prime}\right)f_R^{\prime}%\frac{\partial f_R}{\partial N}
\ .
\label{1.12}
\eea
where the prime denotes derivative with respect to the number of e-folds $N$. Hence, by using Eqs.~(\ref{1.12}) above and assuming a particular Hubble parameter evolution $H=H(N)$, the corresponding Lagrangian $f(R,\xi)$ can be in principle reconstructed. In addition, we are now able to relate the underlying theory and the inflationary parameters, assuming the validity of slow-roll inflation, as will be shown in the upcoming section.

%%%%%%%%%%%%%%%%%%%%%%%%%%%
\section{Slow-roll inflation in $f(R,\xi)$ gravity: inflationary parameters}
\label{Sec:Slow-roll}
%%%%%%%%%%%%%%%%%%%%%%%%%%%

Let us start by briefly reviewing the basic properties of slow-roll inflation by using the most common model, i.e., a single non-minimally coupled scalar field whose Lagrangian is given by
\be
\label{2.1}
S_{\phi} = \int {\rm d}^4 x \sqrt{-g} \left[- \frac{1}{2}\partial_\mu \phi \partial^\mu \phi - V(\phi) \right]\, ,
\ee
By assuming the spatially flat FLRW metric (\ref{1.10}), the equations describing the dynamics of the Universe evolution are
\bea
\frac{3}{\kappa^2} H^2 \Eqn{=} \frac{1}{2}{\dot \phi}^2 + V(\phi)\, ,  \nn
- \frac{1}{\kappa^2} \left( 3 H^2 + 2\dot H \right)
\Eqn{=} \frac{1}{2}{\dot \phi}^2 - V(\phi)\, ,
\label{2.2}
\eea
whereas the scalar field satisfies
\be
\ddot{\phi}+3H\dot{\phi}+\frac{\partial V(\phi)}{\partial\phi}=0
\label{2.3}
\ee
Note that  adequate combinations of the FLRW Eqs.~(\ref{2.2}) together with the assumption of a particular Hubble parameter $H(N)$, easily provide the corresponding inflaton model, as obtained from action (\ref{2.1}), namely \cite{Elizalde:2008yf}
\bea
\omega(\phi) &=& -\frac{2 H'(\phi)}{\kappa^2 H(\phi)}\, , \nn
V(\phi) &=&  \frac{1}{\kappa^2}\left[ 3 \left(H(\phi)\right)^2 + H(\phi) H'(\phi) \right]\, .
\label{2.6}
\eea
Here we have redefined the scalar field as $\phi=N$ and $\omega(\phi)$ refers to the kinetic factor that one needs to introduce to recast the kinetic term as in (\ref{2.1}).
%{\color{red} [The problem is that $\omega(\phi)$ has not been used in the action nor in the EoM]}.
Let us now obtain the corresponding observables in terms of the model (\ref{2.1}). In slow-roll inflation, the scalar field behaves approximately as an effective cosmological constant, since
\be
H\dot{\phi}\gg\ddot{\phi}\,,\quad V\gg\dot{\phi}^2\ .
\label{2.3a}
\ee
This basically means that the friction term in (\ref{2.3}) dominates, the Hubble parameter being approximately constant, $H\sim H_0$. After an enough number of e-folds (around $N=50-65$), the scalar field $\phi$ rolls down the potential slope while the kinetic term of the scalar field increases and eventually dominates over the potential. Thus the field continues oscillating around the minimum of the potential, emitting particles and consequently reheating the Universe. One of the most relevant consequences of inflation is the fast growth of the scalar and tensor fluctuations produced during inflation, which then form the seeds of large-scale structures. Such fluctuations have a characteristic amplitude and scale
dependence which under the slow-roll approximation, are related to the so-called slow-roll parameters defined as follows 
\be
\epsilon=
\frac{1}{2\kappa^2} \left( \frac{V'(\phi)}{V(\phi)} \right)^2\, ,\quad
\eta= \frac{1}{\kappa^2} \frac{V''(\phi)}{V(\phi)}\, , \quad
\lambda^2 = \frac{1}{\kappa^4} \frac{V'(\phi) V'''(\phi)}{\left(V(\phi)\right)^2}\, .
\label{2.4}
\ee
%\ee
%
Here the primes refer to derivatives with respect to the scalar field $\phi$. During inflation, the quantities in (\ref{2.4}) are small enough
in order to expand the required number of e-folds, namely $\epsilon\ll 1$ and $\eta<1$. At the end of  inflation, $\epsilon\gtrsim 1$. Then, for the above inflationary model (\ref{2.1}) the  spectral index $n_\mathrm{s}$ of the curvature perturbations,
the tensor-to-scalar ratio $r$ of the density perturbations,
and the running of the spectral index $\alpha_\mathrm{s}$ can be written in terms of the slow-roll parameters as follows
%
%\be
\be
n_\mathrm{s} - 1= - 6 \epsilon + 2 \eta\, ,\quad r = 16 \epsilon \, , \quad \alpha_\mathrm{s} = \frac{{\rm d} n_\mathrm{s}}{{\rm d}\log k} \sim 16\epsilon \eta - 24 \epsilon^2 - 2 \xi^2\, .
\label{2.5}
\ee
Hence, by using the recent constraints provided by the Planck and Bicep2 collaborations every slow-roll inflationary model can be scrutinised against observational results (see Ref.~\cite{Planck-Inflation}).
%
%
%Note that the aim in this investigation is to provide the above expressions (\ref{2.5}) in terms of the underlying theory, which is expected to be capable of reproducing slow-roll inflation, in order to compare with the observational data. 
Similarly, the slow-roll parameters can be expressed in terms of the Hubble parameter by using Eqs. (\ref{2.6}), which leads to~\cite{Bamba:2014daa} %{\color{red} Maybe the next three expressions in an Appendix?}
%\begin{align}
\begin{eqnarray}
\epsilon &=&- \frac{H(N)}{4 H'(N)} \left[ \frac{6H'(N)}{H(N)}
+ \frac{H''(N)}{H(N)} + \left( \frac{H'(N)}{H(N)} \right)^2 \right]^2
\left( 3 + \frac{H'(N)}{H(N)} \right)^{-2} \, , \nn
\eta &=&-\frac{1}{2} \left( 3 + \frac{H'(N)}{H(N)} \right)^{-1} \left[
\frac{9H'(N)}{H(N)} + \frac{3H''(N)}{H(N)} + \frac{1}{2} \left( \frac{H'(N)}{H(N)} \right)^2
-\frac{1}{2} \left( \frac{H''(N)}{H'(N)} \right)^2
+ \frac{3H''(N)}{H'(N)} \right.\nonumber\\
&&\left.+ \frac{H'''(N)}{H'(N)} \right] \, , \nn
\xi^2 &=&\frac{1}{4} \left( 3 + \frac{H'(N)}{H(N)} \right)^{-2}
\left[\frac{6H'(N)}{H(N)} + \frac{H''(N)}{H(N)}
+ \left( \frac{H'(N)}{H(N)} \right)^2 \right]
\left[ \frac{3H(N) H'''(N)}{H'(N)^2} + \frac{9H'(N)}{H(N)}
\right .
\nonumber \\
\hspace{-4.5mm}
&&\left. +\, \frac{4H''(N)}{H(N)}
+ \frac{5H'''(N)}{H'(N)}-\frac{2H(N) H''(N) H'''(N)}{\left(H'(N)\right)^3} 
+ \frac{H(N) \left(H''(N)\right)^3}{\left(H'(N)\right)^4} 
\right.\nn
\hspace{-4.5mm}
&& \left.
 -\,  \frac{3H(N) \left(H''(N)\right)^2}{\left(H'(N)\right)^3} - \left( \frac{H''(N)}{H'(N)} \right)^2 +\frac{15H''(N)}{H'(N)}+\frac{H(N) H''''(N)}{\left(H'(N)\right)^2} \right]\, .
\label{2.7}
\end{eqnarray}
%*** recordemos \phi and N has been indentified after (3.4) in the rest paper
where let's keep in mind that throughout the paper 
prime shall denote derivative with respect to $\phi\equiv N$.
Turning back to the more general scalar-tensor Lagrangian (\ref{1.1}) and using the equations of motion (\ref{1.12}), the Hubble parameter and its derivatives with respect to the number of e-foldings can be expressed solely
in terms of the  gravitational Lagrangian $f(R,\xi)$ and its derivatives, yielding
\bea
H^2&=&\frac{Rf_R-f}{6\left(f_R+f_R^{\prime}\right)-f_{\xi}\phi^{\prime 2}}\ , \nn
\frac{H^{\prime}}{H}&=&-f_R^{\prime 2}-f_R^{\prime}+\frac{1}{2}f_{\xi}\phi^{\prime 2}}{2f_R+f_R^{\prime}\ ,\nn
\frac{H^{\prime\prime}}{H}&=&...
\label{2.8}
\eea
Here the prime denotes derivatives with respect to the number of e-foldings $N$.  Then, the slow-roll parameters (\ref{2.7}) can be expressed in terms of the Lagrangian $f(R,\xi)$, in analogy to the case of  single field inflation (\ref{2.4}). The explicit expressions of the slow-roll parameters (\ref{2.7}) can be then expressed in terms of the gravitational action (\ref{1.1}) by using (\ref{2.8}). For the sake of clarity, we omit those expressions here. Hence, by following the sketched procedure, the corresponding observational parameters are obtained. In addition, the Lagrangian (\ref{1.1}) can be reconstructed when considering a particular inflationary model which can be fully described by the Hubble parameter $H(N)$. Thus, in the following section we shall consider several inflationary models %characterised by the Hubble parameter 
and study the possibility that they are being driven by
%assuming some %parameterisations
paradigmatic forms of the gravitational Lagrangian (\ref{1.1}).

%%%%%%%%%%%%%%%%%%%%%%%%%%%%%%%%%%%%%%%%%%%%%%%%%%%%%%%%%%%%%%%%%%%%%%%%%%%%%%
\section{Reconstruction and Inflationary constraints of $f(R,\xi)$ gravity} % for some inflationary models}
\label{Sec:Reconstruction}
%%%%%%%%%%%%%%%%%%%%%%%%%%%%%%%%%%%%%%%%%%%%%%%%%%%%%%%%%%%%%%%%%%%%%%%%%%%%%%

%%%%%%%%%%%%%%%%%%%%%%%%%%%%%%%%%%%%%%%%%%%%%%%%%%%%%
In order to reconstruct $f(R,\xi)$ Lagrangians, which are provided by a given cosmological expansion history, two particular classes of models, 
 whose interest shall be explained immediately, will be considered in this Section.
Thus, in the following Sections \ref{SubSec:ModelA}  and \ref{SubSec:ModelB}
we shall consider two paradigmatic examples illustrating the reconstruction procedure. For each of them we shall obtain parameters constraints using slow-roll parameters bounds as provided by Planck and Bicep2 collaborations latest data.

\subsection{$f(R,\xi)=\alpha(\xi)R$ Model}
\label{SubSec:ModelA}
Let us first consider a class of models with an arbitrary function $\alpha(\xi)$ coupled to the standard Einstein-Hilbert gravitational Lagrangian proportional to the Ricci curvature $R$, such that
% $K$-essence model, where the general action is given by,
\be
f(R,\xi)=\alpha(\xi)R\,.
\label{action1}
\ee
Such a model could be motivated, as mentioned in the Introduction, by the scalar field terms emerging when string corrections couple with the effective Einsteinian gravity. Moreover,  provided the coupling $\alpha$ is analytic at $\xi=0$, the Lagrangian (\ref{action1}) can be expanded around $\xi=0$ in order that
\begin{eqnarray}
\label{1}
f(R,\xi)\approx\, \alpha(0)R+ \alpha_{\xi}(0)\xi R + \frac{1}{2}\alpha_{2\xi}(0)\xi^2 R +\mathcal{O}(\xi^3R)
\end{eqnarray}
Thus, such models can be thought of as including an Einstein-Hilbert term -- with a gravitational constant $\alpha(0)$ -- plus non-minimal couplings of the form $\xi^{n} R, \, (n=1,2,...)$ which are indeed expected when for instance $f(R)$ theories are rearranged in the scalar-tensor picture. Moreover, provided the main contribution in $\alpha$ is assumed to be almost linear in $\xi$ the leading term for the non-minimal coupling will be given by $\xi R$, a term which is free of Ostrogradski's instability and that will again be considered in combination with the Starobinsky inflation in Sec.~\ref{Subsection:5.2}.

%
%One can easily understand that,  provided $\alpha$ is analytic at $\xi=0$, such Lagrangian can be expanded as $f(R,\xi)\approx \alpha(0)R+ \alpha_{\xi}(0)\xi R + \frac{1}{2}\alpha_{2\xi}(0)\xi^2 R +\mathcal{O}(\xi^3R)$, so that the leading term for the non-minimal coupling can be thought as $\xi R$.
%

Thus, for the Lagrangian (\ref{action1}), the  FLRW equations (\ref{1.12}) become
\bea
3\alpha (\xi) H^2&=&\frac{R}{2}\alpha_{\xi}(\xi)H^2\phi^{\prime2}-3H^2\frac{\partial \alpha(\xi)}{\partial N}\ , \nn
-H\left(2 H^{\prime}+3H\right)\alpha(\xi)&=&H^2\frac{\partial^2 \alpha(\xi)}{\partial N^2}+H\left(2H+H^{\prime}\right)\frac{\partial \alpha(\xi)}{\partial N}\ .
\label{3.1}
\eea
%{\color{red} If prime is derivative wrt to $N$ I would write $f_R^{\prime}$, etc. in the previous two equations.}\\
%{\color{red} Alvaro: I am lost with primes. Is that derivative wrt $\phi$?}
Note that the second equation in (\ref{3.1}) is actually a differential equation for $\alpha(\xi(N))=\alpha(N)$ which can be solved by specifying the Hubble parameter,
\be
H^2\frac{\partial^2 \alpha(N)}{\partial N^2}+\left(HH^{\prime}+2H^2\right)\frac{\partial \alpha(N)}{\partial N}+\left(2HH^{\prime}+3H^2\right)\alpha(N)=0.
\label{3.2}
\ee
After having solved the equation above for $\alpha(N)$, the corresponding scalar field sector, i.e., the scalar field $\phi$ evolution, can be fully reconstructed using the first FLRW equation in (\ref{3.1}), yielding
\be
\alpha_{\xi}\phi^{\prime 2}=\frac{\alpha+\alpha^{\prime}}{H(2H+H^{\prime})}\ ,
\label{3.2a}
\ee
where one first needs to specify a form of $\alpha(\xi)$.
For instance, for a linear dependence such as $\alpha=\xi\equiv-\frac{1}{2}\partial_{\mu}\phi\partial^{\mu}\phi-V(\phi)$, and after redefining the scalar field such that $\phi=N$, the scalar field sector for this type of actions can be completely reconstructed, 
\be
\omega(\phi)=6\frac{\alpha+\alpha^{\prime}}{6H(2H+H^{\prime})}\ ,\quad
V(\phi)=-\alpha(\phi)-\frac{1}{2}H^2\omega(\phi)\ .
\label{3.2b}
\ee
Note that, here, $H=H(\phi)$ is expressed in terms of the scalar field $\phi$, so that by assuming a Hubble parameter evolution, the full action is reconstructed following the above steps. Let us then consider some inflationary models and reconstruct the corresponding $\alpha(\xi)$. Firstly, we consider the  Hubble expansion \cite{Bamba:2014wda}:
\be
H (N)=\sqrt{c_0 N + c_1}\, ,
\label{3.3}
\ee
where $\{c_0,\, c_1\}$ are constants. Note that slow-roll inflation occurs provided the Hubble parameter  $H_{\mathrm{inf}}$ during inflation can be approximated by a constant as long as both  $c_1/c_0 \gg N$ and $c_1>0$ hold, and eventually decays when $c_0<0$ ({\it c.f.} \cite{Bamba:2014wda} for further details). Then, the equation (\ref{3.2}) turns into
\be
2(c_1+c_0N)\alpha^{\prime\prime}(N)+(c_0+4c_1+4c_0N)\alpha^{\prime}(N)+2(c_0+3c_1+3c_0N)\alpha(N)=0\ .
\label{3.4}
\ee
This equation can be easily solved, leading to the following general solution:
\begin{eqnarray}
\alpha(N)=\e^{-x/c_0}x^{1/4}\left[A_1\mathcal{J}^{1/4}\left(\frac{x}{c_0\sqrt{2}}\right)+A_2\mathcal{Y}^{1/4}\left(\frac{x}{c_0\sqrt{2}}\right)\right]\ ,
\end{eqnarray}
where
%\be
%$x=c_1+c_0N$,
%\label{3.5}
%\ee
$x=c_1+c_0N$,  $\{A_1, A_2\}$ are integration constants and $\{\mathcal{J}^n, \mathcal{Y}^{n}\}$ hold for the Bessel functions of the first and second kind, respectively. Then, the corresponding scalar field sector $\alpha(\xi)$ is obtained following (\ref{3.2a}). In the event of a linear dependence $\alpha\propto\xi$, the kinetic term and scalar potential are given by (\ref{3.2b}).\\

Let us now consider another type of inflationary model, described in this case by an exponential expansion:
\be
H (N)=\sqrt{c_2 \e^{\beta N} + c_3}\, ,
\label{3.6}
\ee
where $c_2$, $c_3$, and $\beta (>0)$ are constants. Note that this type of model mimics a power-law inflation, where
the scale factor is given by $a = \bar{a} t^{\hat{n}}$. It is straightforward to show that during inflation
the Hubble parameter (\ref{3.6}) turns out to be $H^2 = \left( \hat{n}/t \right)^2 = \hat{n}^2\exp\left(-2N/\hat{n}\right)$, where $c_2 \sim \hat{n}^2$, $\beta= -2/\hat{n}$, and $c_3 \sim 0$. Then, as in the previous example, we can reconstruct the gravitational Lagrangian by use of Eq.~(\ref{3.2}), which in this case becomes
\begin{eqnarray}
2x\alpha^{\prime\prime}(x)+\left[x(4+\beta)-c_3\beta\right]\alpha^{\prime}(x)+\left[x(3+\beta)-c_3\beta\right]\alpha(x)=0\ ,
\label{3.7}
\end{eqnarray}
where
%\be
$x=c_3+c_2\e^{\beta N}$\ .
%\label{3.8}
%\ee
Thus, this field equation can be solved, leading to:
\begin{eqnarray}
\alpha(x)&=&\left[A_3\ \mathcal{U}\left(a,b+1,\frac{1}{2}x\sqrt{-32-8\beta+\beta^2}\right)+A_4\ \mathcal{L}\left(a,b,\frac{1}{2}x\sqrt{-32-8\beta+\beta^2}\right)\right]\nonumber\\
&\times&\e^{\frac{-\left(4+\beta+\sqrt{-32-8\beta+\beta^2}\right) x+(2+c_3\beta)\log x}{4}}\nonumber\\ 
%\label{3.9}
\end{eqnarray}
with constants $a$ and $b$ defined as
\be
a=\frac{c_3\beta(\beta-4)-(4+c_3\beta)\sqrt{-32-8\beta+\beta^2}}{4\sqrt{-32-8\beta+\beta^2}}\ , \quad b=\frac{2+c_3\beta}{2}\ .
\label{3.9}
\ee
Here $\mathcal{U}$ is the confluent hypergeometric function and $\mathcal{L}$ is the Laguerre polynomial, whereas $\{A_3, A_4\}$ are integration constants. As in the previous case, the full scalar field sector $\alpha(\xi)$ is obtained once the dependence of $\alpha=\alpha(\xi)$ is found by integrating
Eq.~(\ref{3.2a}). Hence, the above inflationary models can be reproduced within this kind of scalar-tensor theories with non-minimal couplings. In the following section, both models (\ref{3.3}) and (\ref{3.6}) are constrained below by using the Planck/Bicep2 data.\\

\subsection{$f(R,\xi)=\alpha(\xi)+\gamma(R)$ Model}
\label{SubSec:ModelB}
Let us now consider a similar $K$-essence model, whose gravitational Lagrangian is given by,
\be
f(R,\xi)=\alpha(\xi)+\gamma(R)\,,
\label{action2}
\ee
 i.e., the gravitational Lagrangian is split into two different contributions: the first function $\gamma(R)$ corresponds to a standard scalar-tensor $f(R)$ theory, whose physical motivation as next-order terms in the low-energy correction to GR was thoroughly explained in the Introdduction. The second function $\alpha(\xi)$ , i.e.,  a
$K$-essence Lagrangian, may for instance represent the first approximation to either a dark fluid driving a late-time acceleration or, as is the interest of the present manuscript, to a slow-roll inflaton field.
\\

In this case, the FLRW equations (\ref{1.12}) yield
\bea
3H^2\gamma_R&=&\frac{1}{2}\alpha_{\xi}H^2\phi^{\prime 2}+\frac{1}{2}\left(R\gamma_R-\gamma-\alpha\right)-3H^2\frac{\partial \gamma}{\partial N}\ , \nn
-H\left(2 H^{\prime}+3H\right)\gamma_{R}&=&\frac{1}{2}\left(\gamma+\alpha-R\gamma_R\right)+H^2\frac{\partial^2 \gamma_R}{\partial N^2}+H\left(2H+H^{\prime}\right)\frac{\partial \gamma_R}{\partial N}\ .
\label{3.10}
\eea
Here the second equation above provides an algebraic expression for $\alpha(\xi)$:
\be
\alpha=-\gamma-(6H^2+4HH^{\prime}-R)\gamma_R-2\left[(2H^2+H H^{\prime})R^{\prime}+H^2R^{\prime\prime}\right]\gamma_{2R}-2H^2R^{\prime 2}\gamma_{3R}\ .
\label{3.11}
\ee
Then, by the first FLRW equation in (\ref{3.10}) and redefining again the scalar field as $\phi=N$, the corresponding kinetic term is obtained:
\be
\frac{1}{2}\alpha_{\xi}\omega(\phi)=3\gamma_R-\frac{1}{2H^2}\left(R\gamma_R-\gamma-\alpha\right)+3\frac{\partial \gamma}{\partial N}\ ,
\label{3.12}
\ee
%{\color{red} Once again $\omega(\phi)$ was not defined in the original $\xi$.}\\

%{\color{red} ?Es la siguiente frase que yo escribo correcta?}
Thus, for inflationary models of the form (\ref{3.3}) and (\ref{3.6}),  the equations (\ref{3.11}) and (\ref{3.12}) can be solved and the functions  $\gamma(R)$ and $\alpha=\alpha(\xi)$ determined, at least numerically, so that the corresponding action (\ref{action2}) can be reconstructed. In fact, for the paradigmatic inflationary Hubble parameter parameterisations above, analytical expressions are neither obtained for $\gamma(R)$ nor for $\alpha=\alpha(\xi)$ so we decided not to illustrate the whole results, although numerical $\gamma(R)$ and $\alpha=\alpha(\xi)$ could be easily depicted.\\

 In Fig. \ref{fig1}  the classes of models given by Eqs. (\ref{3.3}) and (\ref{3.6}) have been constrained  by using the data released from a joint analysis of the Bicep2/Kerr array and Planck missions \cite{Planck-Inflation}. Here we use the values provided by such analyses, namely 
\be
 n_s=0.968\pm 0.006\ , \quad r< 0.07 \ .
 \label{3.13}
 \ee
 As shown in Fig.~\ref{fig1}, the models studied above can lie within the range provided by the observational data (\ref{3.13}) but depending on the free parameters, they can also present large deviations, particularly in the scalar-to-tensor ratio.
%
%{\color{red} Esto hay que hacerlo aqui, ?verdad?}

\begin{figure}[t]
  \centering
\includegraphics[width=0.4\textwidth]{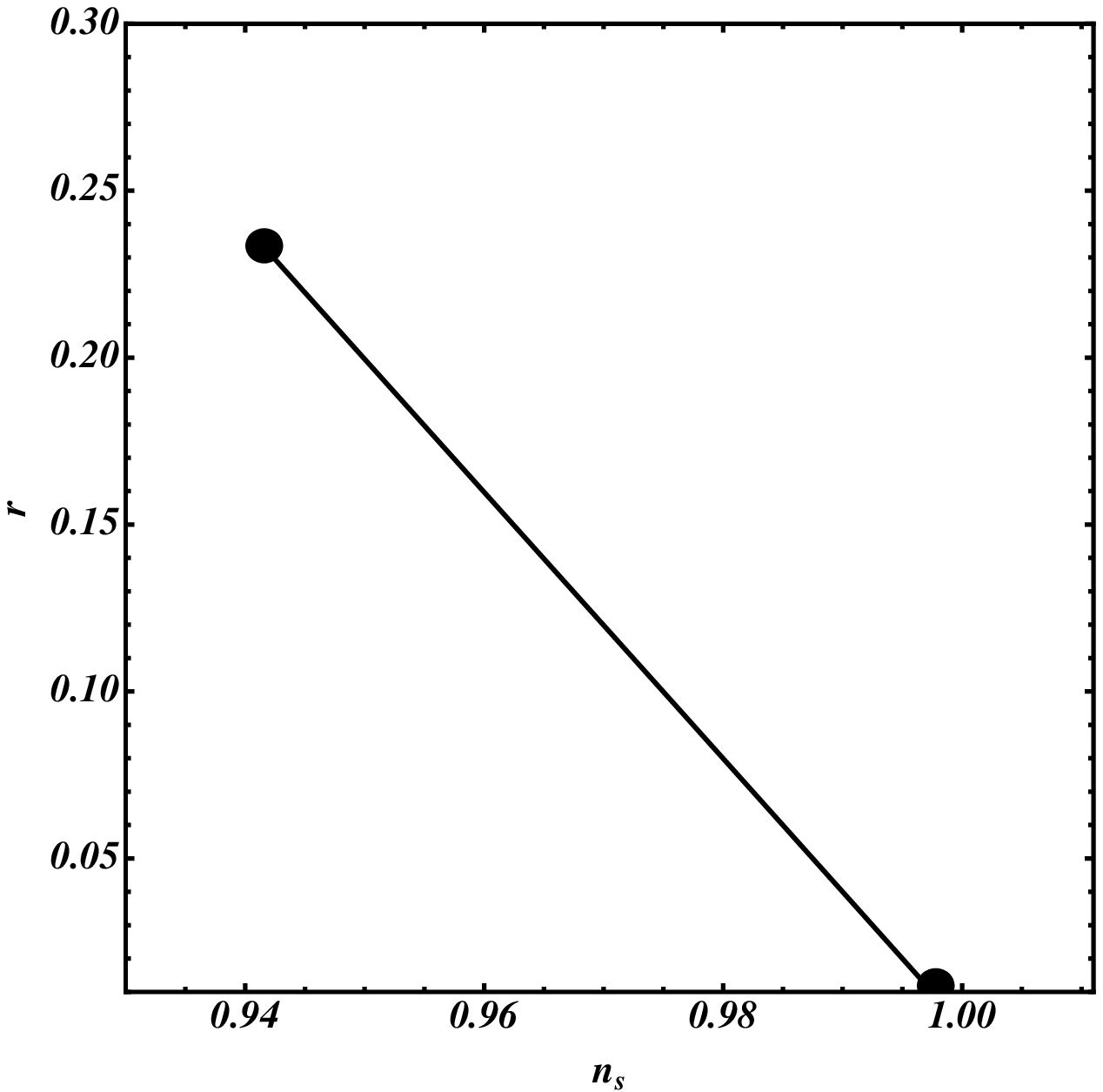}
\includegraphics[width=0.4\textwidth]{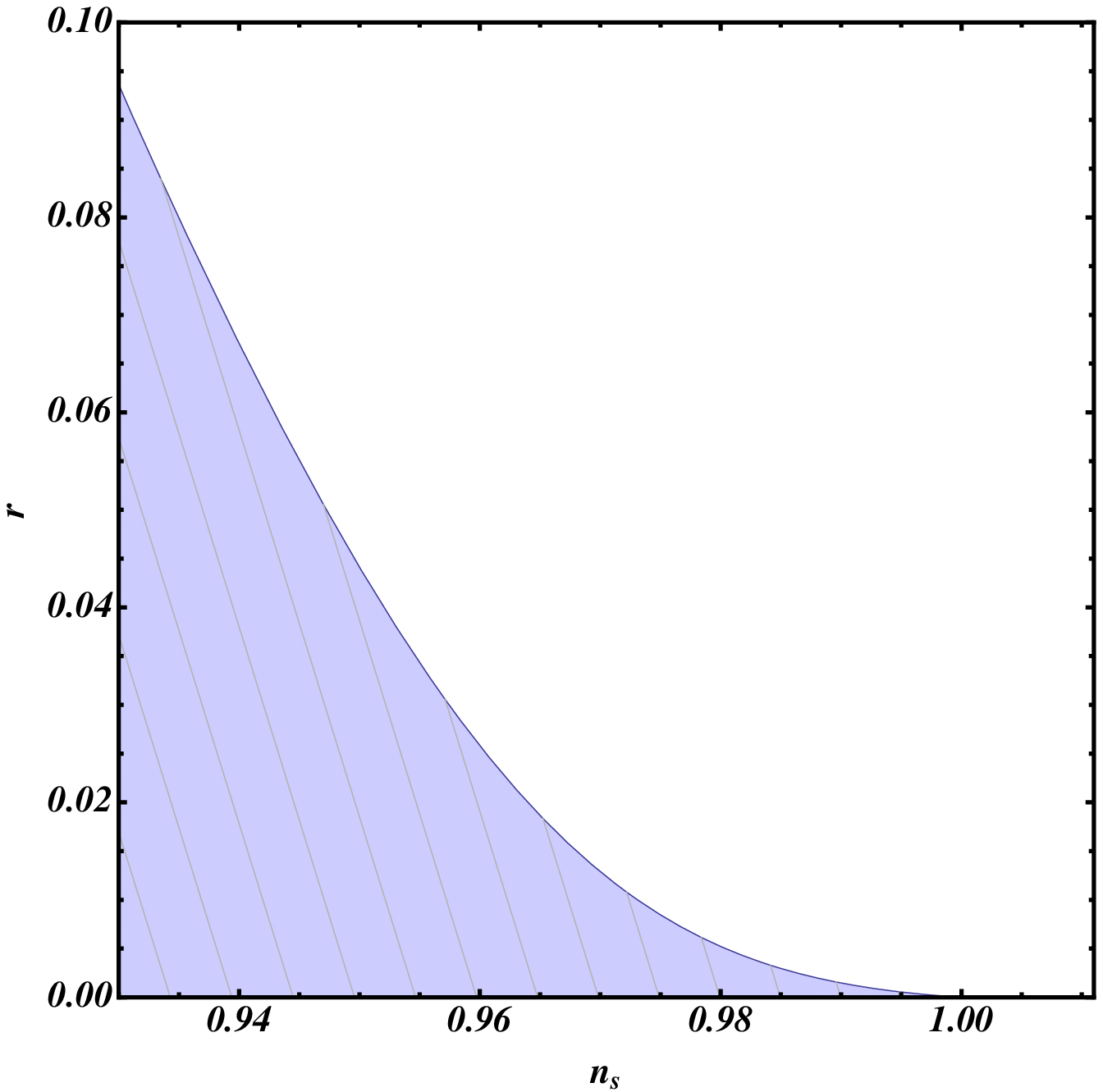}
\caption{
Predictions for the inflationary models given by Eqs. (\ref{3.3}) and (\ref{3.6}). Left panel: the scalar-to-tensor ratio $r_s$ vs. the spectral index $n_s$ for the model (\ref{3.3}), assuming $N=50-65$ and the appropriate range for $c_1/c_0$, namely $c_1/c_0 \gg N$ and $c_1>0$. Right panel: the same for the exponential model (\ref{3.6}). As shown, the model (\ref{3.6}) fits better the observational constraints as provided by Bicep2/Kerr array and Planck.} \label{fig1}
\end{figure}

%%%%%%%%%%%%%%%%%%%%%%%%%%%%%%%%%%%%%%%%%%%%%%%%%%%%%%%%%%%%%%%%%%%%%%%%%%%%%
\section{Deviations from Starobinsky inflation}
\label{Sec:Deviations}
%%%%%%%%%%%%%%%%%%%%%%%%%%%%%%%%%%%%%%%%%%%%%%%%%%%%%%%%%%%%%%%%%%%%%%%%%%%%

Let us now study some deviations from the so-called Starobinsky inflation \cite{staro}, where an action of the kind (\ref{1.1}) is considered. Starobinsky inflation has been a remarkably successful inflationary model, capable of satisfying with great accuracy the observational constraints. The model is given by a particular case of $f(R)$ theories, where a quadratic scalar curvature correction is added to the standard Einstein-Hilbert term, namely
\be
f(R) = R + \frac{R^2}{6\, m^2}\,,
\label{4.1}
\ee
the parameter $m^2$ being a free constant to be constrained by the observations, providing the mass of the dynamical scalar degree of freedom present in the model, inherent to every $f(R)$ model and dubbed {\it scalaron}. In this Section we will analyze models that can be thought of as slight departures from this paradigmatic case. For the sake of simplicity, we perform our calculations in the Einstein frame, where they turn out to be more straightforward. Hence, the action (\ref{4.1}) can be written in as a Brans-Dicke-like action which reads as follows
\be
S=\frac{1}{2\kappa^2}\int {\rm d}^4x\sqrt{-g}\left[\varphi R-Q(\varphi)\right]\ .
\label{4.1a}
\ee
After applying the conformal transformation (\ref{1.5}), the action is shifted to the Einstein frame and becomes %leading to
\be
\tilde{S}=\int {\rm d}^4x\sqrt{-\tilde{g}}\left[\frac{\tilde{R}}{2\kappa^2}-\frac{1}{2}\partial_{\mu}\tilde{\varphi}\partial^{\mu}\tilde{\varphi}-\tilde{Q}(\tilde{\varphi})\right]\ ,
\label{4.1b}
\ee
where the new scalar field and potential are 
\be
\tilde \varphi = \sqrt{\frac{3}{2}} \kappa^{-1} \log \left(1 + \frac{\varphi}{3m^2} \right)\ , \quad \tilde{Q}( \tilde{\varphi}) = \frac{3}{4 \kappa^2} m^2 \left( 1 - \e^{- \sqrt{2/3} \kappa \tilde{\varphi}} \right)^2 .
%\label{V-Star-Einstein}
\label{4.1c}
\ee
Hence, when considered in the Einstein frame, the Starobinsky model (\ref{4.1}) is described by (\ref{4.1b}). If one now  assumes a flat Robertson-Walker geometry in the Einstein frame, which remains a good approximation during inflation, both frames (Einstein vs. Jordan) are connected as follows
\bea
&&{\rm d}s^{2} = -{\rm d}\tilde{t}^2 + \tilde{a}(\tilde{t})^2 {\rm d}{\vec r}^{\,2}\,, \quad \Omega^{-1} {\rm d}\tilde{t}={\rm d}t\,, \quad \Omega^{-1}\tilde{a}(\tilde{t})=a(t)\ ,
\label{4.1d}
\eea
where we recall that the conformal transformation connecting both frames is given by $\Omega^2=\varphi=e^{\sqrt{2/3} \kappa \tilde{\varphi}}$. Hence, the FLRW equations to be obtained from Eq. (\ref{4.1b})
correspond to inflationary models with a single scalar field minimally coupled, i.e., Eqs. (\ref{2.2}), such that slow--roll inflation occurs in the regime $\kappa \phi \gg 1$, i.e., as long as the field $\tilde{\phi}$ rolls down the flat part of the potential (\ref{4.1c}) slowly enough, and the kinetic energy of the inflaton remains much smaller than its potential energy, $\dot{\tilde{\phi}}^2 \ll \tilde{Q}(\tilde{\phi})$. Then the Friedman equation (\ref{2.2}) together with the scalar field equation become, approximately,
$\tilde{H}^2  \simeq  \frac{\kappa^2}{6}  \tilde{Q}(\tilde{\phi})$, and $3\tilde{H} \dot{\tilde{\phi}} \simeq -  \tilde{Q}'(\tilde{\phi})$. The slow--roll parameters (\ref{2.4}) read, in this case,
\bea
\epsilon &\simeq& \frac{1}{2 \kappa^2} \left( \frac{\tilde{Q}'(\tilde{\varphi})}{\tilde{Q}(\tilde{\varphi})} \right)^2\; \simeq \frac{4}{3} \left(\e^{\sqrt{\frac{2}{3}}\kappa \tilde{\varphi}} -1 \right)^{-2}  \simeq \frac{4}{3} \e^{-2\sqrt{\frac{2}{3}}\kappa \tilde{\phi}}\ , \\ % \nonumber
\eta &\simeq& \frac{1}{\kappa^2}\frac{\tilde{Q}''(\tilde{\varphi})}{\tilde{Q}(\tilde{\varphi})}\; \simeq \frac{4}{3} \frac{2-\e^{\sqrt{\frac{2}{3}}\kappa \tilde{\varphi}}}{\left(-1+\e^{\sqrt{\frac{2}{3}}\kappa \tilde{\varphi}}\right)^{2}}  \simeq -\frac{4}{3} \e^{-\sqrt{\frac{2}{3}}\kappa \tilde{\varphi}}.
\eea
In the last step of the above relations we have assumed that $\kappa \tilde{\varphi} \gg 1$. Inflation ends provided that $\epsilon \simeq 1$, which corresponds to the inflaton kinetic energy becoming important. As usual, the amount of inflation is measured by the number of e-foldings $N$, which is defined as
\begin{align}
N  \equiv  \int_{t_{start}}^{t_{end}} \tilde{H} {\rm d}t.
\label{N_Staro}
\end{align}
To reach agreement with observations, one usually requires that $N \simeq 55 - 65$. Now, under the slow--roll approximation the above relation (\ref{N_Staro}) takes the following approximate form
\begin{align}
N &\simeq -\kappa^2 \int_{\tilde{\phi}_{start}}^{\tilde{\varphi}_{end}} \frac{\tilde{Q}(\tilde{\varphi})}{\tilde{Q}'(\tilde{\varphi})} {\rm d}\varphi
\simeq \frac{3}{4}e^{\sqrt{2/3}\kappa \; \tilde{\varphi}_{start}}.
\end{align}
%%%%
In the previous expression we assumed again that $\kappa \tilde{\varphi} \gg 1$, in the slow--roll approximation, and the integration is performed with respect to $\tilde{\varphi}$. Notice that the number of e-foldings is related to the slow-roll parameters, as
\be
\epsilon \simeq \frac{3}{4}\frac{1}{N^2}\ , \quad \eta \simeq -\frac{1}{N}\ .
\label{SRparamStaro}
\ee
% The scalar and tensor fluctuations produced during inflation have a characteristic amplitude and scale dependence, which in the Einstein frame, under the slow-roll approximation, are given by the standard relations for GR minimally coupled to a scalar field \cite{Mukhanov},
% \begin{align}
% & P_{\mathcal{R}} \simeq  \frac{1}{4\pi^2}\frac{\tilde{H}^4}{\dot{\tilde{\phi}}^2 } \simeq \frac{N^2}{3 \pi}  \frac{m^2}{m_{p}^2}, \\
% & P_{GW} \simeq \frac{16}{\pi} \frac{\tilde{H}^2}{m_{p}^2} \simeq \frac{4}{ \pi}  \frac{m^2}{m_{p}^2},
% \end{align}
% while the tensor to scalar ratio follows from the above relations as $r \equiv P_{GW} /P_{\mathcal{R}}\simeq  (48/\pi)N^{-2} \simeq 16 \epsilon$. Notice that in the final approximate equalities of above relations we restricted ourselves to the Starobinsky potential for $\kappa \tilde{\phi} \gg 1$, under the slow-roll approximation.
The spectral indices corresponding to the scalar and tensor fluctuations, under the slow-roll regime are respectively given by the relations (\ref{2.5}).
% \begin{align}
% & n_{S} - 1 \simeq -4 \epsilon - 2\eta, \\
% & n_T \simeq - 16\epsilon,
% \end{align}
% with $\eta$ the second slow--roll parameter defined as $\eta \equiv \ddot{\tilde{\phi}}/(\tilde{H} \dot{\tilde{\phi}})$.\\
Remind that from the joint analysis of the BICEP2/Keck Array and Planck data, one gets the values given in (\ref{3.13}).
%a value for the spectral index of $n_s=0.968\pm 0.006$, and an upper bound for the scalar-to-tensor ratio $r< 0.07$ comes from a joint analysis of the BICEP2/Keck Array and Planck data \cite{Planck-Inflation}. 
In order to illustrate the power of Starobinsky inflation,  the assumption of a number of e-folds $N_e=50-65$, leads to the following values of the inflationary observables:
%%%%%%%%
\be
n_s=0.958-0.968\ , \quad r=0.004-0.002\,,
\label{SpRStaro}
\ee
in agreement with observations, particularly when assuming a number of e-foldings to lie near $65$.
In order to analyse certain modifications of the Starobinsky inflation, two different actions extending such an inflationary model will be considered. They are the following

\subsection{Minimally coupled $K$-essence}
\label{Subsection:5.1}
Let us here consider an action of the type (\ref{action2}),
\be
S=\frac{1}{2\kappa^2}\int {\rm d}x^4\sqrt{-g}\left[R + \frac{R^2}{6 m^2}+2\kappa^2\alpha(\xi)\right]\ .
\label{action3}
\ee
In this action, the coupling between the added scalar field in the form of a $K$-essence theory and the curvature terms  -- Einstein-Hilbert plus the paradigmatic quadratic Starobinsky term -- is minimal. Thus the eventual deviations observed  for such a model can indeed be solely attributed to the presence of the extra field $\alpha(\xi)$ in the action, whereas the somewhat desirable minimal coupling between curvature and matter will be preserved. Moreover, this model could be considered as the simplest expected departure from the Starobinsky inflation, i.e., the one given by a minimally coupled scalar field. Finally, as we shall illustrate in what follows, the model (\ref{action3}) can in fact be reinterpreted as an inflationary scenario with two scalar fields. This approach has been considered, although from different points of view, in the existing literature mentioned in the Introduction.

This action, similarly to (\ref{1.3}), can be rewritten in terms of only scalar fields. Nevertheless, in the previous action the scalar field sector and the curvature are not directly coupled, so that only one auxiliary field is required. The introduction of such an auxiliary field leads (\ref{action3}) to
\be
S=\frac{1}{2\kappa^2}\int {\rm d}^4x\sqrt{-g}\left[\varphi R-Q(\varphi)+2\kappa^2\alpha(\xi)\right]\ .
\label{4.2}
\ee
At this stage, the action is transformed to the Einstein frame by means of the conformal transformation (\ref{1.5}), what yields
\be
\tilde{S}=\int {\rm d}^4x\sqrt{-\tilde{g}}\left[\frac{\tilde{R}}{2\kappa^2}-\frac{1}{2}\partial_{\mu}\tilde{\varphi}\partial^{\mu}\tilde{\varphi}-\tilde{Q}(\tilde{\varphi})+\e^{-2\sqrt{\frac{2}{3}}\kappa\tilde{\varphi}}\alpha\left(\tilde{\xi}\right)\right]\ .
\label{4.3}
\ee
Recall  $\tilde{\xi}=-\frac{1}{2}\e^{\sqrt{\frac{2}{3}}\kappa\tilde{\varphi}}\partial_{\mu}\phi\partial^{\mu}\phi-V(\phi)$ and note that the scalar field $\tilde\varphi$ and its self-interacting term $\tilde Q(\tilde\varphi)$ are given by expressions in (\ref{4.1c}).
% \be
% \tilde \varphi = \sqrt{\frac{3}{2}} \kappa^{-1} \log \left(1 + \frac{\varphi}{3m^2} \right)\ , \quad \tilde{Q}( \tilde{\varphi}) = \frac{3}{4 \kappa^2} m^2 \left( 1 - e^{- \sqrt{2/3} \kappa \tilde{\varphi}} \right)^2 . \label{V-Star-Einstein}
% \label{4.4}
% \ee
% Hence, the action in the Einstein-frame action is described by the action (\ref{4.4}). Assuming a flat FLRW metric in the Einstein frame, the FLRW metric (\ref{1.10}) in the Jordan frame is connected with the Eintein frame as follows
% \bea
% ds^{2} = -d\tilde{t}^2 + \tilde{a}(\tilde{t})^2 d{\bf x}^2\ ,\nn
% \Omega^{-1} d\tilde{t}=dt\ \quad \Omega^{-1}\tilde{a}(\tilde{t})=a(t)\ ,
% \label{4.5}
% \eea
Thus, the FLRW equations in this spacetime, analogously to Eqs. (\ref{2.2}), are given by  %{\color{red} Is not there any evolution equation for $\xi$?}
\bea
\tilde{H}^2 \equiv \left( \frac{\dot{\tilde{a}}}{\tilde{a}} \right)^2 &=& \frac{\kappa^2}{3}\left[\frac{ \dot{\tilde{\varphi}}^2 }{2 } +  \tilde{Q}(\tilde{\varphi})+\e^{-\sqrt{\frac{2}{3}}\kappa\tilde{\varphi}}\alpha_{\xi}\dot{\phi}^2-\e^{-2\sqrt{\frac{2}{3}}\kappa\tilde{\varphi}}\alpha(\tilde{\xi})\right], \nn
-3\tilde{H}^2-2\dot{\tilde{H}}&=&\kappa^2\left[\frac{ \dot{\tilde{\varphi}}^2 }{2 } -  \tilde{Q}(\tilde{\varphi})+\e^{-2\sqrt{\frac{2}{3}}\kappa\tilde{\varphi}}\alpha(\tilde{\xi})\right]\ .
\label{4.6}
\eea
 After, for the sake of simplicity, the choice
$\alpha(\xi)=\xi=-\frac{1}{2}\partial_{\mu}\phi\partial^{\mu}\phi-V(\phi)$ has been made, the equations for the scalar fields $\tilde\varphi$ and $\phi$ become %such that the scalar field equations read
\bea
\ddot{\tilde{\varphi}}+3H\dot{\tilde{\varphi}}+\tilde{Q}_{,\tilde{\varphi}}&=&\sqrt{\frac{2\kappa^2}{3}}\e^{-\sqrt{\frac{2}{3}}\kappa\tilde{\varphi}}\left[-\frac{1}{2}\dot{\phi}^2+2\e^{-\sqrt{\frac{2}{3}}\kappa\tilde{\varphi}}V(\phi)\right]\ , \nn
\ddot{\phi}+3H\dot{\phi}+V_{,\phi}\e^{-\sqrt{\frac{2}{3}}\kappa\tilde{\varphi}}&=&\sqrt{\frac{2\kappa^2}{3}}\dot{\phi}\dot{\varphi}\ ,
\label{4.7}
\eea
respectively. As a consequence, in this scenario %and assuming that both fields slow roll,
the slow-roll conditions can be expressed as 
\be
\{\dot{\tilde{\varphi}}^2, \dot{\phi}^2\}\ll U(\phi,\tilde{\varphi})\ , \quad \ddot{\tilde{\varphi}}\ll H\dot{\tilde{\varphi}}\ , \quad \ddot{\phi}\ll H\dot{\phi}\ ,
\label{4.8}
\ee
where $U(\phi,\tilde{\varphi})=\tilde{Q}(\tilde{\varphi})+\e^{-2\sqrt{\frac{2}{3}}\kappa\tilde{\varphi}}V(\phi)$. By assuming the slow-roll conditions, the first FLRW equation and the scalar field equations yield
\bea
\frac{3}{\kappa^2}H^2&\simeq& \tilde{Q}(\tilde{\varphi})+\e^{-2\sqrt{\frac{2}{3}}\kappa\tilde{\varphi}}V(\phi)\ ,\nn
3H\dot{\tilde{\varphi}}&\simeq& -\tilde{Q}_{,\tilde{\varphi}}+2\sqrt{\frac{2}{3}}\kappa\e^{-2\sqrt{\frac{2}{3}}\kappa\tilde{\varphi}}V(\phi)\ , \nn
3H\dot{\phi}&\simeq& -V_{,\phi}\e^{-\sqrt{\frac{2}{3}}\kappa\tilde{\varphi}}
\label{4.9}
\eea
Nevertheless, note that in general multifield inflationary models, perturbations not only induce adiabatic modes but also isocurvature fluctuations, so that the curvature perturbation $\mathcal{R}$ is not conserved outside the horizon \cite{GarciaBellido:1995qq}-\cite{Starobinsky:2001xq}:
\be
\dot{\mathcal{R}}=\frac{H}{\dot{H}}\frac{k^2}{a^2}\Phi+C\left(\frac{\delta\tilde{\varphi}}{\dot{\tilde{\varphi}}}-\frac{\delta\phi}{\dot{\phi}}\right)\ .
\label{4.10}
\ee
Here $\Phi$ is the Bardeen potential, and $\{\delta\tilde{\varphi}, \delta\phi\}$ are the fluctuations of the fields. Then, in the long-wavelenght limit $k\rightarrow 0$, the r.h.s. of (\ref{4.10}) does not vanish, in general, unless $C$, which depends on the background evolution, does. The entropy fluctuations are then given by
\be
\delta S\propto \left(\frac{\delta\tilde{\varphi}}{\dot{\tilde{\varphi}}}-\frac{\delta\phi}{\dot{\phi}}\right),
\label{4.11}
\ee
which depends on the trajectories in the scalar field space. Nevertheless, if the scalar field space is not curved, so that both fields behave similarly, the isocurvature perturbations become negligible, such that the spectral index is given by \cite{Sasaki:1995aw}
\be
n_s-1=-6\epsilon+2\eta_{\sigma\sigma}
\label{4.12}
\ee
where
\be
\epsilon=-\frac{\dot{H}}{H^2}\ ,\quad \eta_{\sigma\sigma}=\frac{1}{\kappa^2}\frac{\tilde{\sigma}^{I}\tilde{\sigma}_{J}}{U}\left(\mathcal{G}^{IK}(D_{J}D_{K}U)-\mathcal{R}^{I}_{KLJ}\dot{\sigma}^{K}\dot{\sigma}^{L}\right)\ .
\label{4.13}
\ee
Here $\sigma^{I}=\{\tilde{\varphi},\phi\}$ and $\mathcal{G}_{IJ}$ is the metric of the scalar field space, while $\tilde{\sigma}^{J}=\frac{\sigma^{J}}{\sqrt{\mathcal{G}_{IJ}\dot{\sigma}^{I}\dot{\sigma}^{J}}}$. Note that the expressions (\ref{4.13}) reduces to the corresponding ones for a single scalar field (\ref{2.4}) when we drop out one of the fields. In order to study small deviations from Starobinsky inflation, let us consider both fields to roll down together, so that we can neglect the isocurvature perturbations and compute just the spectral index (\ref{4.12}) when the effects of the extra scalar field are small. For the case of the action (\ref{4.3}) and assuming $\alpha(\xi)=\xi=-\frac{1}{2}\partial_{\mu}\phi\partial^{\mu}\phi-V(\phi)$, the metric of the scalar field space is given by
\be
\mathcal{G}_{IJ}\dot{\sigma}^{I}\dot{\sigma}^{J}=\dot{\tilde{\varphi}}^2+\dot{\phi}^2\e^{-\sqrt{\frac{2}{3}}\kappa\tilde{\varphi}}\ .
\label{4.14}
\ee
Then, from (\ref{4.13}), the slow-roll parameters read
\bea
\epsilon=\frac{1}{2\kappa^2}\frac{\left(\tilde{Q}_{,\tilde{\varphi}}-2\sqrt{\frac{2}{3}}\kappa\e^{-2\sqrt{2/3}\kappa\tilde{\varphi}}V\right)^2+\e^{-3\sqrt{2/3}\kappa\tilde{\varphi}}V_{,\phi}^2}{\left(\tilde{Q}+\e^{-2\sqrt{2/3}\kappa\tilde{\varphi}}V\right)^2}\ , \nn
\eta_{\sigma\sigma}=\frac{1}{\kappa^2}\frac{1}{\dot{\tilde{\varphi}}^2+\dot{\phi}^2\e^{-\sqrt{\frac{2}{3}}\kappa\tilde{\varphi}}}\left(\frac{\dot{\tilde{\varphi}}^2}{U}\frac{\partial^2U}{\partial\tilde{\varphi}^2}+\frac{\dot{\phi}^2}{U}\frac{\partial^2U}{\partial\phi^2}+\frac{\dot{\tilde{\varphi}}\dot{\phi}}{U}\frac{\partial^2U}{\partial\tilde{\varphi}\partial\phi}\right)\ .
\label{4.15}
\eea
Here recall that $U(\phi,\tilde{\varphi})=\tilde{Q}(\tilde{\tilde{\varphi}})+\e^{-2\sqrt{\frac{2}{3}}\kappa\tilde{\varphi}}V(\phi)$. The number of e-foldings during inflation is expressed in terms of the values of the scalar fields as follows:
\begin{eqnarray}
N=\int^{t_{end}}_{t_{start}}H {\rm d}t&=&-\kappa^2\int_{\tilde{\varphi}_{start}}^{\tilde{\varphi}_{end}} \frac{\tilde{Q}(\tilde{\varphi})+\e^{-2\sqrt{\frac{2}{3}}\kappa\tilde{\varphi}}V(\phi)}{\tilde{Q}_{,\tilde{\varphi}}-2\sqrt{\frac{2}{3}}\kappa\e^{-2\sqrt{2/3}\kappa\tilde{\varphi}}V}{\rm d}\tilde{\varphi}\nonumber\\
&=&-\kappa^2\int_{\phi_{start}}^{\phi_{end}} \frac{\tilde{Q}(\tilde{\varphi})+\e^{-2\sqrt{\frac{2}{3}}\kappa\tilde{\varphi}}V(\phi)}{\e^{-\sqrt{2/3}\kappa\tilde{\varphi}}V_{,\phi}}{\rm d}\phi .
\label{4.16}
\end{eqnarray}
As shown, in general it is very difficult to obtain, analytically, information for the modified Starobinsky inflation model (\ref{4.3}), when one considers a generic scalar potential $V(\phi)$. Nevertheless, since we are assuming the scalar field $\phi$ to behave similarly to $\tilde{\varphi}$, in order to avoid isocurvature perturbations, we are here allowed to consider a potential $V(\phi)$ similar to $\tilde{Q}(\tilde{\varphi})$ in (\ref{4.1c}), thus given by
\be
V(\phi)= V_0\left(1-\e^{-k\phi}\right)^2\ ,
\label{4.17}
\ee
where $V_0$ and $k$ are positive constants. Considering now that during inflation
\be
\kappa\tilde{\varphi}\gg 1\ \quad k\phi\gg 1\ ,
\label{4.18}
\ee
the following approximation holds
\be
\tilde{Q}(\tilde{\tilde{\varphi}})+\e^{-2\sqrt{\frac{2}{3}}\kappa\tilde{\varphi}}V(\phi)\sim \tilde{Q}(\tilde{\tilde{\varphi}}),
\label{4.19}
\ee
and the slow-roll parameter (\ref{4.15}) becomes
\bea
\epsilon&\sim&\frac{1}{2\kappa^2}\left(\frac{\tilde{Q}_{,\tilde{\varphi}}}{\tilde{Q}}\right)^2 \simeq \frac{4}{3}\e^{-2\sqrt{\frac{2}{3}}\kappa \tilde{\phi}}\nn
\eta_{\sigma\sigma}&\sim&\frac{1}{\kappa^2}\frac{\dot{\tilde{\varphi}}^2}{\dot{\tilde{\varphi}}^2+\e^{-\sqrt{\frac{2}{3}}\kappa\tilde{\varphi}}\dot{\phi}^2}\frac{\tilde{Q}_{\tilde{\varphi}\tilde{\varphi}}}{\tilde{Q}}\sim\frac{1}{\kappa^2}\frac{\tilde{Q}_{\tilde{\varphi}\tilde{\varphi}}}{\tilde{Q}} \simeq -\frac{4}{3} \e^{-\sqrt{\frac{2}{3}}\kappa \tilde{\varphi}}.
\label{4.20}
\eea
While for the number of e-foldings (\ref{4.16}), we get
\be
N=-\kappa^2\int \frac{\tilde{Q}(\tilde{\varphi})+\e^{-2\sqrt{\frac{2}{3}}\kappa\tilde{\varphi}}V(\phi)}{\tilde{Q}_{,\tilde{\varphi}}-2\sqrt{\frac{2}{3}}\kappa\e^{-2\sqrt{2/3}\kappa\tilde{\varphi}}V}{\rm d}\tilde{\varphi}\sim -\kappa^2\int \frac{\tilde{Q}(\tilde{\varphi})}{\tilde{Q}_{,\tilde{\varphi}}}{\rm d}\tilde{\varphi}\simeq\frac{3}{4}\e^{\sqrt{2/3}\kappa \; \tilde{\varphi}_{start}}\ .
\label{4.21}
\ee
In this way, the predictions from Starobinsky inflation (\ref{SpRStaro}) are fully recovered. As expected, the term in front of the scalar potential $V(\phi)$, in the equations above, screens this potential, so that no deviation is induced, as long as one considers a potential of the form (\ref{4.17}). However, any other choice that may compensate the screening will lead to curves in the scalar field space, thereby inducing isocurvature modes and also providing deviations in the adiabatic modes.\\

\subsection{Non-minimally coupled $K$-essence}
\label{Subsection:5.2}
A second class of modification %of Starobinsby inflation
can be described by the following action
\be
S=\frac{1}{2\kappa^2}\int {\rm d}^4x\sqrt{-g}\left[(R+\xi) + \frac{\left(R+\xi\right)^2}{6 m^2}\right]\ .
\label{4.22}
\ee
Such a choice, apart from adding a linear term in $\xi$ to the Starobinsky model,  also hosts
a non-minimal coupling between the Ricci curvature and the extra scalar field of the form $\xi\,R$, which is indeed free of the Ostrogradski instability when the action is mapped into 
the Einstein frame, and a quadratic term $\xi^2$ which can be understood as a $\mathcal{L}_2$ Horndeski-like term, or in other words, is nothing but a $K$-essence contribution. Thus, model 
(\ref{4.22}) is extending the Starobinsky inflation with a scalar field as well as allowing for the existence of a non-minimal coupling between this field and the geometric curvature in the aim of unveiling the effect that such coupling may have when compared to previous results in Section~\ref{Subsection:5.2}.

Then, following the procedure described in Sec. \ref{Sec:f(R,xi)}, the action above can be rewritten by using an auxiliary field $\varphi$ as follows:
\be
S=\frac{1}{2\kappa^2}\int {\rm d}^4x\sqrt{-g}\left[\varphi(R+\xi)-Q(\varphi)+2\kappa^2\mathcal{L}_m\right]\ ,
\label{4.23}
\ee 
where the potential is given by,
\be
Q(\varphi)=\frac{3m^2}{2}\left(\varphi-1\right)^2\ .
\label{4.24}
\ee
As already carefully explained, by applying the conformal transformation (\ref{1.5}), the action is transformed into the Einstein frame, leading to
\be
\tilde{S}=\int {\rm d}^4x\sqrt{-\tilde{g}}\left[\frac{\tilde{R}}{2\kappa^2}-\frac{1}{2}\partial_{\mu}\tilde{\varphi}\partial^{\mu}\tilde{\varphi}-\tilde{Q}(\tilde{\varphi})-\frac{1}{2}\partial_{\mu}\tilde{\phi}\partial^{\mu}\tilde{\phi}-\e^{-\sqrt{\frac{2}{3}}\kappa\tilde{\varphi}}\tilde{V}\left(\tilde{\phi}\right)\right]\ ,
\label{4.25}
\ee
where $\tilde{\phi}=\phi/\sqrt{2\kappa^2}$ and $\tilde{V}=V/2\kappa^2$, while the potential is transformed as 
\be
\tilde{Q}( \tilde{\varphi}) = \frac{3}{4 \kappa^2} m^2 \left( 1 - e^{- \sqrt{2/3} \kappa \tilde{\varphi}} \right)^2\,,
\ee
 keeping the same form as in usual Starobinsky inflation.  The FLRW equation and the equations for both scalar fields are given by  
\bea
&&\tilde{H}^2 = \frac{\kappa^2}{3}\left[\frac{ \dot{\tilde{\varphi}}^2 }{2 } +  \tilde{Q}(\tilde{\varphi})+\frac{\dot{\phi}^2}{2}+\e^{-\sqrt{\frac{2}{3}}\kappa\tilde{\varphi}}\tilde{V}(\tilde{\phi})\right], \nn
&&\ddot{\tilde{\varphi}}+3H\dot{\tilde{\varphi}}+\tilde{Q}_{,\tilde{\varphi}}-\sqrt{\frac{2\kappa^2}{3}}\e^{-\sqrt{\frac{2}{3}}\kappa\tilde{\varphi}}V(\tilde{\phi})=0\ , \nn
&&\ddot{\tilde{\phi}}+3H\dot{\tilde{\phi}}+\tilde{V}_{,\tilde{\phi}}\e^{-\sqrt{\frac{2}{3}}\kappa\tilde{\varphi}}=0\ ,
\label{4.26}
\eea
 Then, by applying the slow-roll approximations (\ref{4.8}), the equations (\ref{4.26}) turn out:
\bea
\frac{3}{\kappa^2}H^2&\simeq& \tilde{Q}(\tilde{\varphi})+\e^{-2\sqrt{\frac{2}{3}}\kappa\tilde{\varphi}}\tilde{V}(\tilde{\phi})\ ,\nn
3H\dot{\tilde{\varphi}}&\simeq& -\tilde{Q}_{,\tilde{\varphi}}+\sqrt{\frac{2}{3}}\kappa\e^{-\sqrt{\frac{2}{3}}\kappa\tilde{\varphi}}\tilde{V}(\tilde{\phi})\ , \nn
3H\dot{\tilde{\phi}}&\simeq& -\tilde{V}_{,\tilde{\phi}}\,\e^{-\sqrt{\frac{2}{3}}\kappa\tilde{\varphi}}
\label{4.27}
\eea
From (\ref{4.13}), the slow-roll parameters read
\bea
\epsilon=\frac{1}{2\kappa^2}\frac{\left(\tilde{Q}_{,\tilde{\varphi}}-\sqrt{\frac{2}{3}}\kappa\e^{-\sqrt{2/3}\kappa\tilde{\varphi}}\tilde{V}\right)^2+\e^{-2\sqrt{2/3}\kappa\tilde{\varphi}}\tilde{V}_{,\tilde{\phi}}^{\,2}}{\left(\tilde{Q}+\e^{-\sqrt{2/3}\kappa\tilde{\varphi}}\tilde{V}\right)^2}\ , \nn
\eta_{\sigma\sigma}=\frac{1}{\kappa^2}\frac{1}{\left(\dot{\tilde{\varphi}}^2+\dot{\tilde{\phi}}^2\right)U}\left(\dot{\tilde{\varphi}}^2\frac{\partial^2U}{\partial\tilde{\varphi}^2}+\dot{\tilde{\phi}}^2\frac{\partial^2U}{\partial\tilde{\phi}^2}+\dot{\tilde{\varphi}}\dot{\tilde{\phi}}\frac{\partial^2U}{\partial\tilde{\varphi}\partial\tilde{\phi}}\right)\ .
\label{4.28}
\eea
Here $U=\tilde{Q}(\tilde{\varphi})+\e^{-\sqrt{2/3}\kappa\tilde{\varphi}}\tilde{V}$. While the number of e-foldings can be expressed in terms of the scalar fields as:
\be
N=\int^{t_{end}}_{t_{start}}H {\rm d}t=-\kappa^2\int_{\tilde{\varphi}_{start}}^{\tilde{\varphi}_{end}} \frac{\tilde{Q}(\tilde{\varphi})+\e^{-\sqrt{\frac{2}{3}}\kappa\tilde{\varphi}}\tilde{V}(\tilde{\phi})}{\tilde{Q}_{,\tilde{\varphi}}-\sqrt{\frac{2}{3}}\kappa\e^{-\sqrt{2/3}\kappa\tilde{\varphi}}\tilde{V}}{\rm d}\tilde{\varphi}=-\kappa^2\int_{\tilde{\phi}_{start}}^{\tilde{\phi}_{end}} \frac{\tilde{Q}(\tilde{\varphi})+\e^{-\sqrt{\frac{2}{3}}\kappa\tilde{\varphi}}\tilde{V}(\tilde{\phi})}{\e^{-\sqrt{2/3}\kappa\tilde{\varphi}}\tilde{V}_{,\tilde{\phi}}}{\rm d}\tilde{\phi} .
\label{4.29}
\ee
As above, we may consider the scalar potential $\tilde{V}(\tilde{\phi})=V_0\left(1-\e^{-k\tilde{\phi}}\right)^2$, and assume that both fields slow roll similarly in order to avoid isocurvature modes. By a first look, such assumptions, together with (\ref{4.18}) would lead to similar results as above, i.e. to the Starobinsky predictions. However, since the kinetic term of $\tilde{\phi}$ is not screened by the other scalar field as in the case above, this leads to $\dot{\tilde{\varphi}}^2+\dot{\tilde{\phi}}^2\sim2\dot{\tilde{\varphi}}^2$, and the slow-roll parameters can be approximated as follows:
\bea
\epsilon&\sim&\frac{1}{2\kappa^2}\left(\frac{\tilde{Q}_{,\tilde{\varphi}}}{\tilde{Q}}\right)^2 \simeq \frac{4}{3} e^{-2\sqrt{\frac{2}{3}}\kappa \tilde{\phi}}\nn
\eta_{\sigma\sigma}&\sim&\frac{1}{\kappa^2}\frac{\dot{\tilde{\varphi}}^2}{\dot{\tilde{\varphi}}^2+\dot{\tilde{\phi}}^2}\frac{\tilde{Q}_{\tilde{\varphi}\tilde{\varphi}}}{\tilde{Q}}\sim\frac{1}{2\kappa^2}\frac{\tilde{Q}_{\tilde{\varphi}\tilde{\varphi}}}{\tilde{Q}} \simeq -\frac{4}{6} e^{-\sqrt{\frac{2}{3}}\kappa \tilde{\varphi}}.
\label{4.30}
\eea
And the number of e-foldings leads to,
\be
N\simeq\frac{3}{4}e^{\sqrt{2/3}\kappa \; \tilde{\varphi}_{start}}\ .
\label{4.31}
\ee
Then, the slow-roll parameters (\ref{4.31}) can be expressed in terms of the number of e-foldings:
\be
\epsilon \simeq \frac{3}{4}\frac{1}{N^2}\ , \quad \eta_{\sigma\sigma} \simeq -\frac{1}{2N}\ .
\label{4.32}
\ee
Hence, the factor $1/2$ does not affect the scalar-to-tensor ratio $r=16\epsilon$ but it does the spectral index. Then, in order to satisfy the Planck constraints, inflation has to be shorter, i.e. the number of e-foldings has to be smaller. Whether one assumes $N\sim50-65$, the following spectral index and scalar-to-tensor ratio are obtained:
\be
n_s=0.978-0.983\ \quad r=0.004-0.002\ .
\label{4.33}
\ee
It must be noted that while the scalar-to-tensor ratio satisfies the Planck/Bicep2 constraint of $r<0.07$, this is not the case for the spectral index, whose value provided by Planck is $n_s=0.968\pm 0.006$, such that model (\ref{4.22}) predicts certain values that lie outside of the $1\sigma$ region provided by Planck, independent of the number of e-foldings, that are indeed closer to scale-invariant perturbations. In addition, other choices of the scalar potential $\tilde{V}(\tilde{\phi})$ will likely lead to isocurvature perturbations, so that this model has an slightly different predictions than Starobinsky inflation, although the values (\ref{4.33}) may be still considered as viable. \\

Other choices of the initial conditions for the scalar fields would also lead to curves in the field space, i.e., both scalar fields will roll down differently and non-adiabatic modes will appear, largely deviating from the Starobinsky inflation. In order to illustrate such a case, let us consider the model (\ref{4.22}), but where scalar fields roll down differently. We may assume that $\dot{\tilde{\varphi}}=n\dot{\tilde{\phi}}$, where $n$ is a real number. Hence, the slow-roll parameters (\ref{4.13}) are now given by,
\bea
\epsilon&\sim&\frac{1}{2\kappa^2}\left(\frac{\tilde{Q}_{,\tilde{\varphi}}}{\tilde{Q}}\right)^2 \simeq \frac{4}{3} e^{-2\sqrt{\frac{2}{3}}\kappa \tilde{\phi}}\nn
\eta_{\sigma\sigma}&\sim&\frac{1}{\kappa^2}\frac{\dot{\tilde{\varphi}}^2}{\dot{\tilde{\varphi}}^2+\dot{\tilde{\phi}}^2}\frac{\tilde{Q}_{\tilde{\varphi}\tilde{\varphi}}}{\tilde{Q}}\sim\frac{n}{(n+1)\kappa^2}\frac{\tilde{Q}_{\tilde{\varphi}\tilde{\varphi}}}{\tilde{Q}} \simeq -\frac{n}{n+1} e^{-\sqrt{\frac{2}{3}}\kappa \tilde{\varphi}}.
\label{4.34}
\eea
While the number of e-foldings can be well approximated by (\ref{4.31}). Then, the slow-roll parameters lead to:
\be
\epsilon \simeq \frac{3}{4 N^2}\ , \quad \eta_{\sigma\sigma} \simeq -\frac{n}{(n+1)N}\ .
\label{4.35}
\ee
Therefore, the differences in the scalar field trajectories only affect the parameter $\eta_{\sigma\sigma}$ and consequently the spectral index. Besides the contribution of the isocurvature modes that likely appear in this case, we can easily obtain the spectral index and the scalar-to-tensor ratio (\ref{4.12}) by considering  the case where $\tilde{\varphi}$ rolls down much faster $(n=100)$. Then, assuming a number of e-foldings $N=50-65$ once again, the following values are obtained:
\be
n_s=0.958-0.968\ \quad r=0.004-0.002\ .
\label{4.36}
\ee
As shown, the spectral index may be closer to the Planck value (and consequently to the pure Starobinsky inflation) than in the case where both fields roll down similarly (\ref{4.33}). However, whether one considers $\tilde{\phi}$ to roll down much faster $(n=1/100)$, this leads to
\be
n_s=0.997-0.998\ \quad r=0.004-0.002\ ,
\label{4.37}
\ee
whose value of the spectral index provides quasi-scale invariant perturbations, deviating from the Starobinsky model. Nevertheless, in the case of both fields rolling down differently, the appearance of isocurvature modes would modify the above expressions and would increase the difference between the models.

% Instead of considering particular choice of $V(\phi)$, different than (\ref{4.17}), let us express the slow-roll parameters for the Starobinsky model (\ref{SRparamStaro}) with a correction:
%\be
%\epsilon \simeq \frac{3}{4}\frac{1}{N^2}+\lambda N+O(N^2)\ , \quad \eta \simeq -\frac{1}{N}+\lambda N+O(N^2)\ .
%\label{4.22}
%\ee

% The Hubble function $\tilde{H}$ is approximately constant during inflation, and its time variation is described by the slow--roll parameter $\epsilon$ defined as
% \begin{align}
% \epsilon & \equiv - \frac{\dot{\tilde{H}}}{\tilde{H}^2}.
% \end{align}

\section{Conclusions}
\label{Sec:Conclusions}

In this paper we have explored the inflationary paradigm in the framework of theories beyond General Relativity where non-minimal couplings of the scalar curvature with a scalar field % Lagrangian 
are considered. As we have proved in Sec. \ref{Sec:Reconstruction}, inflation can certainly be achieved in these classes of action, which extend both the usual $K$-essence theories and the standard scalaron field. %minimimally coupled to the Einstein-Hilbert action. 
Thus we have also shown in the same Section that all the observational constraints provided by the Planck survey latest data  \cite{Planck-Inflation} are smoothly satisfied. In this sense, we have developed a reconstructing method for the gravitational action capable of reproducing convenient slow-roll inflationary models. Then, once the expressions for the slow-roll parameters for a general evolution are set, the corresponding constraints for every model are obtained. As one of the main results herein, a wide range of inflationary models were proved 
to fulfill such constraints and, consequently, the class of theories analysed giving rise to such inflationary models can be reconstructed. \\

Moreover, we have then considered in Sec.~\ref{Sec:Deviations} the so-called Starobinsky inflation to which an additional scalar field is added. There we have analysed two different ways of modifying Starobinsky inflation. In both of them, we have shown that similar results as in ordinary Starobinsky inflation can be achieved as well as provided the conditions on the extra scalar field potential to be satisfied. In this sense, when one assumes that both fields slow roll similarly, the isocurvature perturbations become negligible, whereas the adiabatic ones lead approximately to the same result as in standard Starobinsky inflation. Nevertheless, we have seen that any other choice for the potential will lead to corrections that may, in principle, be kept under control by the appropriate parameters, but which will inevitably give rise to a non-negligible isocurvature mode, which may rule out the model. In addition, whilst the first modification Starobinsky inflation as given in (\ref{action3}) leads to the same predictions as the original Starobinsky model, provided the above assumptions are considered, the second model (\ref{4.22}) analysed here yields a slightly different value of the spectral index, which do not lie within the error bars provided by Planck \cite{Planck-Inflation}. Generally, we can say that, as in previously considered multifield inflationary models, one needs 
to impose strong conditions on the extra fields in order to guarantee its range of validity. Thus, deviations as the ones considered here turn out to be severely constrained. \\

In conclusion, we have carefully analysed $f(R,\xi)$ theories and their viability in order to provide realistic inflationary scenarios, which is clearly achieved in some specific cases. In addition, we have shown that modifications of Starobinsky inflation are allowed, as far as the extra fields slow roll in a similar fashion as the standard inflaton does and we have shown that those models where the extra scalar field is minimally coupled to the Starobinsky gravitational action are favoured.

\begin{acknowledgments}
%%%%%%%%%%%%%%%%%%%%%%%%%%%%%%%%%%%%%%%%%%%%%
%
A.d.l.C.D. acknowledges financial support from the University of Cape Town (UCT) Launching Grants
programme, National Research Foundation grant 99077 2016-2018, Ref. No. CSUR150628121624, MINECO (Spain) projects FIS2014-52837-P, FPA2014-53375-C2-1-P, Consolider-Ingenio
MULTIDARK CSD2009-00064 and CSIC I-LINK1019 and  would also like to thank the Instittuto de F\'isica Te\'orica (IFT UAM-CSIC, Madrid Spain) for its support via the KA107 action of the Erasmus$+$ Call for international Mobility.

E.E. and S.D.O. are supported in part by MINECO (Spain), Project FIS2013-44881-P, by the CSIC I-LINK 1019 Project, and by the CPAN Consolider Ingenio Project.
D.S.-G. acknowledges support from a postdoctoral fellowship Ref.~SFRH/BPD/95939/2013 by Funda\c{c}\~ao para a Ci\^encia e a Tecnologia (FCT, Portugal) and the support through the research grant UID/FIS/04434/2013 (FCT, Portugal). D.S.-G. also acknowledges the hospitality of Institute of Space Sciences (CSIC-IEEC, Barcelona) during the realisation of this work, funded by CSIC I-LINK1019 Project.

\end{acknowledgments}

\end{document}